\documentclass[final,5p,twocolumn]{elsarticle}

\pdfoutput=1

\makeatletter
\def\ps@pprintTitle{%
	\let\@oddhead\@empty
	\let\@evenhead\@empty
	\def\@oddfoot{}%
	\let\@evenfoot\@oddfoot}
\makeatother

\usepackage{amssymb}
\usepackage{array}
\newcolumntype{M}[1]{>{\centering\arraybackslash}m{#1}}
\usepackage{hyperref}
\usepackage{xcolor}
\hypersetup{
	colorlinks,
	linkcolor={red!50!black},
	citecolor={blue!50!black},
	urlcolor={blue!80!black}
}

\journal{Information Fusion}

\begin{document}

\begin{frontmatter}

\title{GPS-Denied Navigation Using Low-Cost Inertial Sensors and Recurrent Neural Networks\tnoteref{license}}

\tnotetext[license]{\textcopyright 2021. This manuscript version is made available under the CC-BY-NC-ND 4.0 license \url {https://creativecommons.org/licenses/by-nc-nd/4.0/}}

\author[1]{Ahmed AbdulMajuid\corref{cor1}}
\ead{amajuid@hotmail.com}

\author[1]{Osama Mohamady}
\ead{o.mohamady@eng.cu.edu.eg}

\author[1]{Mohannad Draz}
\ead{m_draz@cu.edu.eg}

\author[1]{Gamal El-bayoumi}
\ead{gelbayoumi@cu.edu.eg}

\cortext[cor1]{Corresponding author}

\affiliation[1]{organization={Aerospace Engineering Department, Cairo University},
				addressline={ElGamaa St.}, 
				city={Giza},
				citysep={},
				postcode={12613}, 
				country={Egypt}}
			
\begin{abstract}
Autonomous missions of drones require continuous and reliable estimates
for the drone's attitude, velocity, and position. Traditionally, these states are estimated by applying Extended Kalman Filter (EKF) to Accelerometer, Gyroscope, Barometer, Magnetometer, and GPS measurements. When the GPS signal is
lost, position and velocity estimates deteriorate quickly, especially
when using low-cost inertial sensors. This paper proposes an estimation
method that uses a Recurrent Neural Network (RNN)
to allow reliable estimation of a drone's position and velocity in
the absence of GPS signal. The RNN is trained on a public dataset
collected using Pixhawk. This low-cost commercial autopilot logs
the raw sensor measurements (network inputs) and corresponding
EKF estimates (ground truth outputs). The dataset is comprised of
548 different flight logs with flight durations ranging from 4 to
32 minutes. For training, 465 flights are used, totaling 45 hours.
The remaining 83 flights totaling 8 hours are held out for validation.
Error in a single flight is taken to be the \emph{maximum} absolute
difference in 3D position (MPE) between the RNN predictions (without
GPS) and the ground truth (EKF with GPS). On the validation set, the
median MPE is 35 meters. MPE values as low as 2.7 meters in a 5-minutes
flight could be achieved using the proposed method. The MPE in 90\%
of the validation flights is bounded below 166 meters. The network
was experimentally tested and worked in real-time.
\end{abstract}


\end{frontmatter}

\section{Introduction}

Many applications require accurate positioning in the absence of GPS.
Examples include indoor navigation performed by robots in warehouses
or garages, underwater operations performed by Autonomous Under Water
Vehicles (AUVs), and self-driving cars or drones moving in tunnels,
under bridges, or in dense urban environments.

Many approaches were taken to solve the navigation problem in GPS-denied
environments. In robotics and self-driving cars, it
is common to assist the inertial sensors with cameras \cite{Bloesch2015},
laser scanners \cite{Chiang2019}, radar \cite{Vu2011}, or car On-Board
Diagnostics (OBD) \cite{Choi2017}. However, aside from the added cost,
these sensors impose constraints on the environment or vehicle operation
to function properly.

Before GPS, the classical aerospace positioning method was the Inertial
Navigation System (INS). INS utilizes thoroughly calibrated high-grade
inertial sensors and complex navigation algorithms to estimate
the position from acceleration and angular rates \cite{DavidTitterton2009}. However, the cost of such an approach is not justifiable in a commercial context. 

A newer trend is to use Artificial Intelligence (AI) methods to assist
the INS. These methods can both reduce the need for costly calibration \cite{Chiang2003}
and allow the use of lower-cost inertial sensors \cite{Chiang2009}.

Different AI algorithms can be utilized in various steps in the
inertial navigation process. One popular algorithm is Radial Basis
Function Neural Network (RBFNN), which was used to predict the errors
in position and velocity estimated by the INS given a window of estimates
\cite{Semeniuk2006} or Accelerometer measurements \cite{Shen2019}.
It can also be used along with an EKF to estimate position and attitude
from the Accelerometer, Gyroscope (collectively Inertial Measurement
Unit IMU), and Magnetometer measurements \cite{Wu2019}.

Variations of Multi-Layer Perceptrons (MLP) are also commonly used.
MLPs were used to predict position increments from velocity and heading
\cite{ElSheimy2006}, and predict the errors in heading and velocity
given their changes from one time step to another \cite{wang2006neural}.
They were also used to predict the changes in latitude and longitude
given the IMU measurements and INS velocity \cite{Yao2017}. 

Recently, Recurrent Neural Networks (RNN) are being utilized for their
superiority with time series problems. RNNs were trained to predict 
the INS position and velocity errors using the INS change in velocity
and heading as input \cite{Dai2020}. Long Short-Term Memory (LSTM)
networks, which are enhanced RNNs, can also use the history of Kalman
gain from the Kalman Filter to predict the INS velocity
and position errors \cite{Shen2021}. 

Other methods like Input-Delay Neural Networks (IDNN) \cite{Noureldin2011},
Ensemble Learning \cite{Li2017}, Nonlinear Autoregressive with exogenous
input (NARX) \cite{Li2019,Bai2018}, and Fuzzy Inference Systems (FIS)
\cite{Havyarimana2018} are also applied with promising results.

\subsection*{Contributions in this Paper}
\begin{itemize}
	\item None of the work pointed above used an aerial vehicle as a host; using
	an aerial vehicle introduces more vector components in position, velocity,
	and attitude. Aerial vehicles are also capable of performing more
	complex maneuvers than cars or boats.
	\item Most of the previous work used relatively high-grade sensors (cost 2000\$
	to 9000\$), which are easier to model. This paper utilizes extremely
	low-cost sensors (< 50\$). Low-cost sensors usually have more sources
	of errors and exhibit non-consistent error behaviors. 
	\item The experiments carried in earlier work use the same hardware to collect
	both training and validation data. That is, a single sensor is used
	to collect data in a single long trip using the same host vehicle.
	The work presented in this paper uses data collected from many different
	sensors and host vehicles under different conditions. The error characteristics
	of each sensor unit differ from the others even if all these units
	are of the same sensor model.
\end{itemize}

\section{Pixhawk and PX4 Drone Flight Stack}

The Pixhawk is a commercial low-cost (about \$170) flight controller
board suitable for research and industrial applications. It integrates
an ARM processor, IMU, Barometer, Magnetometer, and additional components
required for flight monitoring and control \cite{Meier2011}. It also
accepts a micro SD card, on which it saves the flight data, including
raw sensor measurements, control actions, hardware status, and other important information. The community that developed the first
Pixhawk is PX4; they maintain a fully functional and open source autopilot
software called PX4 autopilot \cite{Meier2015}. In addition to the
low-level sensor drivers, control loops, and planning algorithm, PX4
autopilot features an INS/GPS integration routine, called EKF2 \cite{Garcia2020},
as a part of its Estimation and Control Library (ECL) \cite{Riseborough2016}.
EKF2 fuses measurements from IMU, Barometer, Magnetometer, and GPS
to estimate the drone's states, namely, attitude represented in quaternions,
velocity, and position in local frame (North-East-Down). EKF2 estimates,
too, are logged on the SD card.

\begin{figure}[h]
	\centering
	\includegraphics[width=1\linewidth]{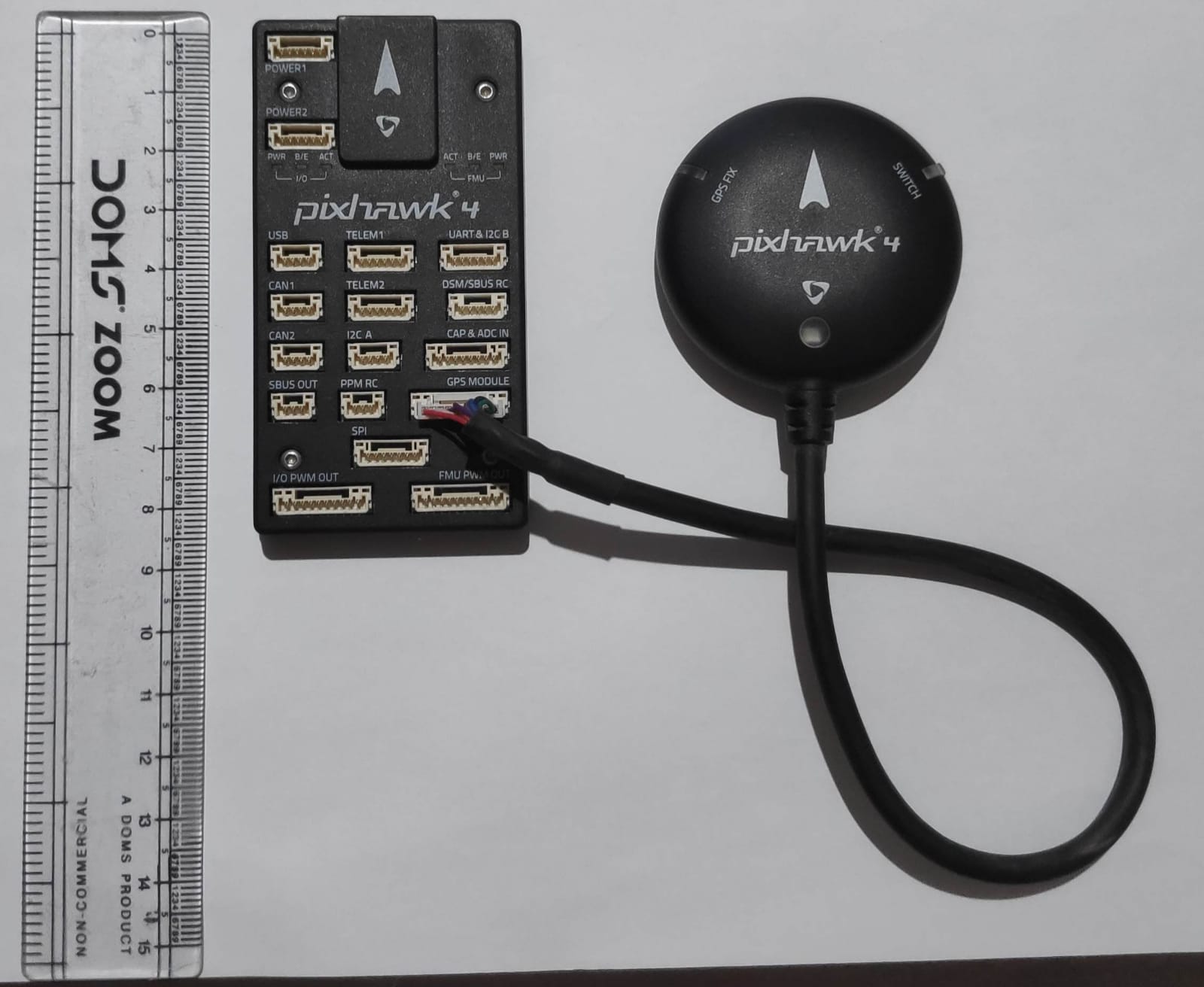}
	\caption{Pixhawk4 autopilot hardware with a Neo M8N GPS Module connected}
	\label{fig:pixhawk4andgps}
\end{figure}

The PX4 team maintains a database of flight logs \cite{PX4.Community2020},
where users worldwide upload actual flight data collected
from real flights. The database contains thousands of logs collected
using different Pixhawk versions and using various host vehicles.
The data used in this work are downloaded from the PX4 database, no
simulated data is used. Only flight logs for Pixhawk4 \cite{Willee2018}
- the most recent Pixhawk version - were downloaded. Figure \ref{fig:pixhawk4andgps}
shows the Pixhawk4 board and the standard M8N GPS used with it. Sensors
models used in the Pixhawk4 board are listed in Table \ref{tab:pixhawk4sensors}.

\begin{table}[h]
	\centering
	\caption{Pixhawk4 sensors}
	\begin{tabular}{|c|c|c|}
		\hline
		Sensor & Model & Price \\
		\hline
		\hline
		IMU & ICM-20689 \cite{Invensense2018} \& BMI055 \cite{Sensortec2014} & \$3 \\
		\hline
		Magnetometer & IST8310 \cite{Isentek2017} & \$3 \\
		\hline
		Barometer & MS5611 \cite{TE2017} & \$10 \\
		\hline
	\end{tabular}
	\label{tab:pixhawk4sensors}
\end{table}

The logs downloaded from the PX4 database contain the raw sensor measurements
(IMU, Barometer, and Magnetometer); those are treated as network inputs
(features). The logs also contain the flight states estimated by the
EKF2 (attitude, velocity, and position); those are treated as network
outputs (labels). Other hardware health data are also logged, and those
are used to remove the corrupted logs. 

Flights used for this work ranged from 4 to 32 minutes, with an average
flight duration of 6 minutes. A total of 548 flights are used, with
a combined duration of about 54 hours. Only 465 flights are used for
training (about 45 hours), and the remaining 83 flights (about 8 hours)
are held for validation. 

\begin{figure}[h]
	\centering
	\includegraphics[width=1\linewidth]{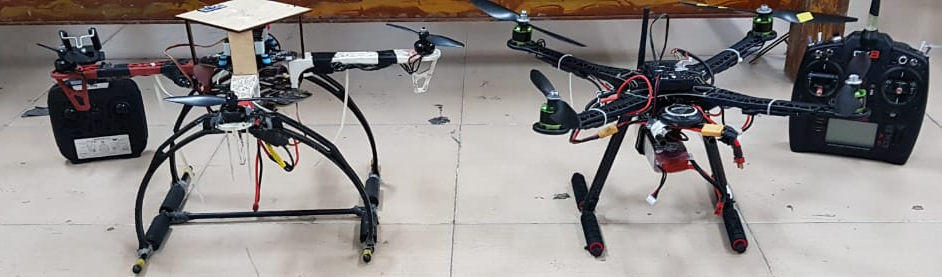}
	\caption{DJI Flamewheel F450 (left) and Holybro S500}
	\label{fig:quads}
\end{figure}

The logs are collected using different types of host vehicles, although
mostly Quadcopters, some Fixed Wing (FW) and Vertical Takeoff and
Landing (VTOL) vehicles were used. Figure \ref{tab:different_host_vehicle}
summarizes different vehicle types found on the database and used
for this work. The most popular vehicles found in the PX4 logs database
are DJI Flamewheel F450 and Holybro S500, shown in Table \ref{fig:quads}

\begin{table}[h]
	\caption{Different host vehicle types}
	\centering{}
	\begin{tabular}{|c|c|}
		\hline 
		Vehicle Type & Number of Logs\\
		\hline 
		\hline 
		Quadrotor & 526\\
		\hline 
		Fixed Wing & 19\\
		\hline 
		Standard VTOL & 8\\
		\hline 
		Octorotor & 4\\
		\hline 
		Tiltrotor VTOL & 2\\
		\hline 
		Hexarotor & 1\\
		\hline 
	\end{tabular}
	\label{tab:different_host_vehicle}
\end{table}

\section{PX4's Extended Kalman Filter Algorithm (EKF2)}

The PX4's EKF2 algorithm consists of two steps; prediction, using
the IMU and inertial navigation techniques, and correction using any
other available sensors. In the prediction step, the angular rates
$\omega_{x},\omega_{y},\omega_{z}$ measured by the Gyroscope in the
vehicle's body frame and the accelerations $a_{x},a_{y},a_{z}$ measured
by the Accelerometer in the same frame are integrated numerically
to obtain delta angles and delta velocities. Then, bias terms are
subtracted, these terms represent all of the inertial navigation error
sources, and as a part of the EKF state vector, they are updated with
each iteration.

\begin{figure*}[h]
	\includegraphics[width=1\textwidth]{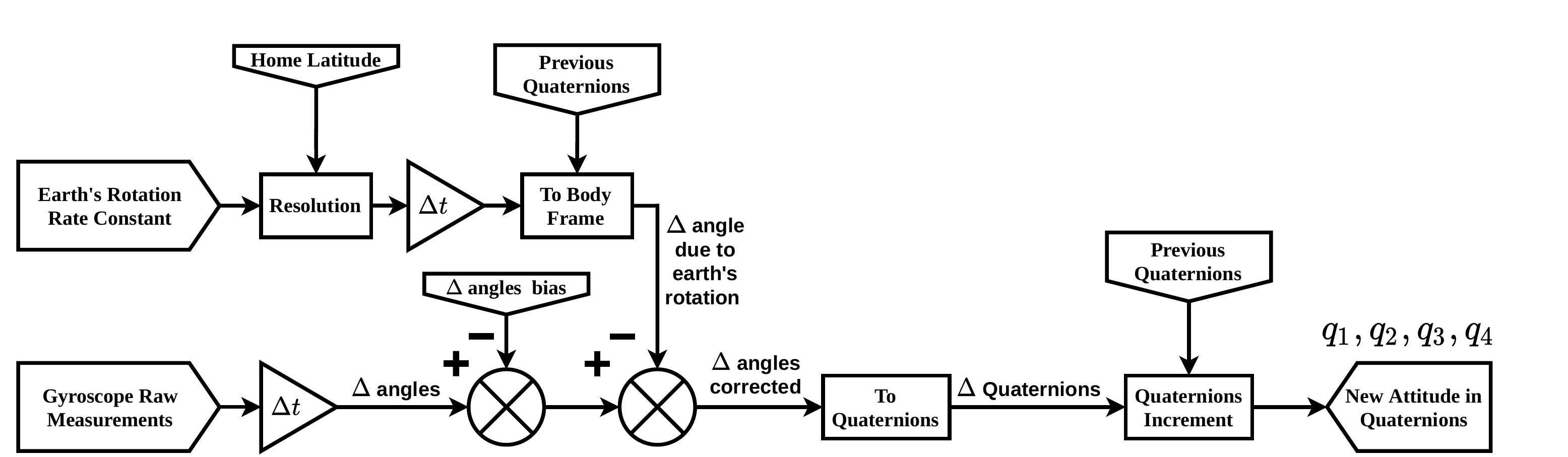}
	\caption{Attitude prediction from Gyroscope measurements in the EKF2 prediction step}
	\label{fig:inertial_nav_gyro}
\end{figure*}

The debiased delta angles are then corrected for the earth's rotation
rate effect, which is a function of the latitude; this can be taken
as the home latitude. The corrected delta angles are transformed to
delta quaternions and those are used with the previous quaternions
to calculate the vehicle's new attitude in quaternions $q_{1},q_{2},q_{3},q_{4}$.
This process is outlined in Figure\ref{fig:inertial_nav_gyro}. 

\begin{figure*}[h]
	\includegraphics[width=1\textwidth]{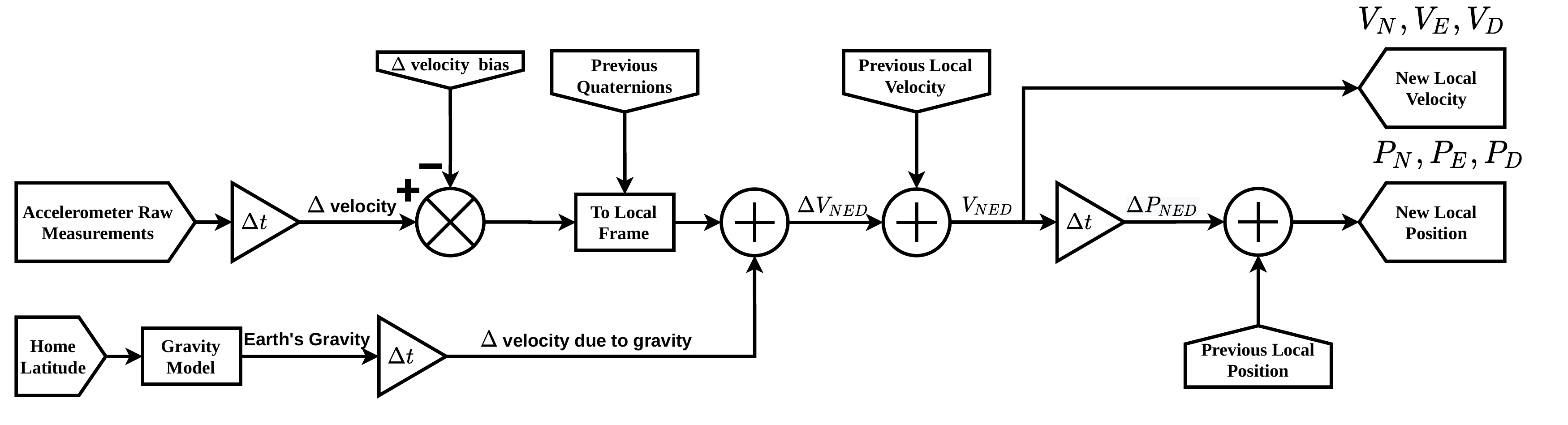}
	\caption{Velocity and position prediction from attitude and Accelerometer measurements in the EKF2 prediction step}
	\label{fig:inertial_nav_accel}
\end{figure*}

This new attitude is then used to transform the debiased delta velocities
from the body frame to the local frame. Then the gravity effect is
added, which is also dependent on the latitude. The resulting corrected
delta velocity is added to the previous velocity to obtain the new
velocity in the local frame $V_{N},V_{E},V_{D}$. Which is integrated
again to obtain the position increments, those are added to the previous
position to obtain the new position in the local frame $P_{N},P_{E},P_{D}$.
This process is outlined in Figure \ref{fig:inertial_nav_accel}.

\begin{figure}[h]
	\centering
	\includegraphics[width=1\linewidth]{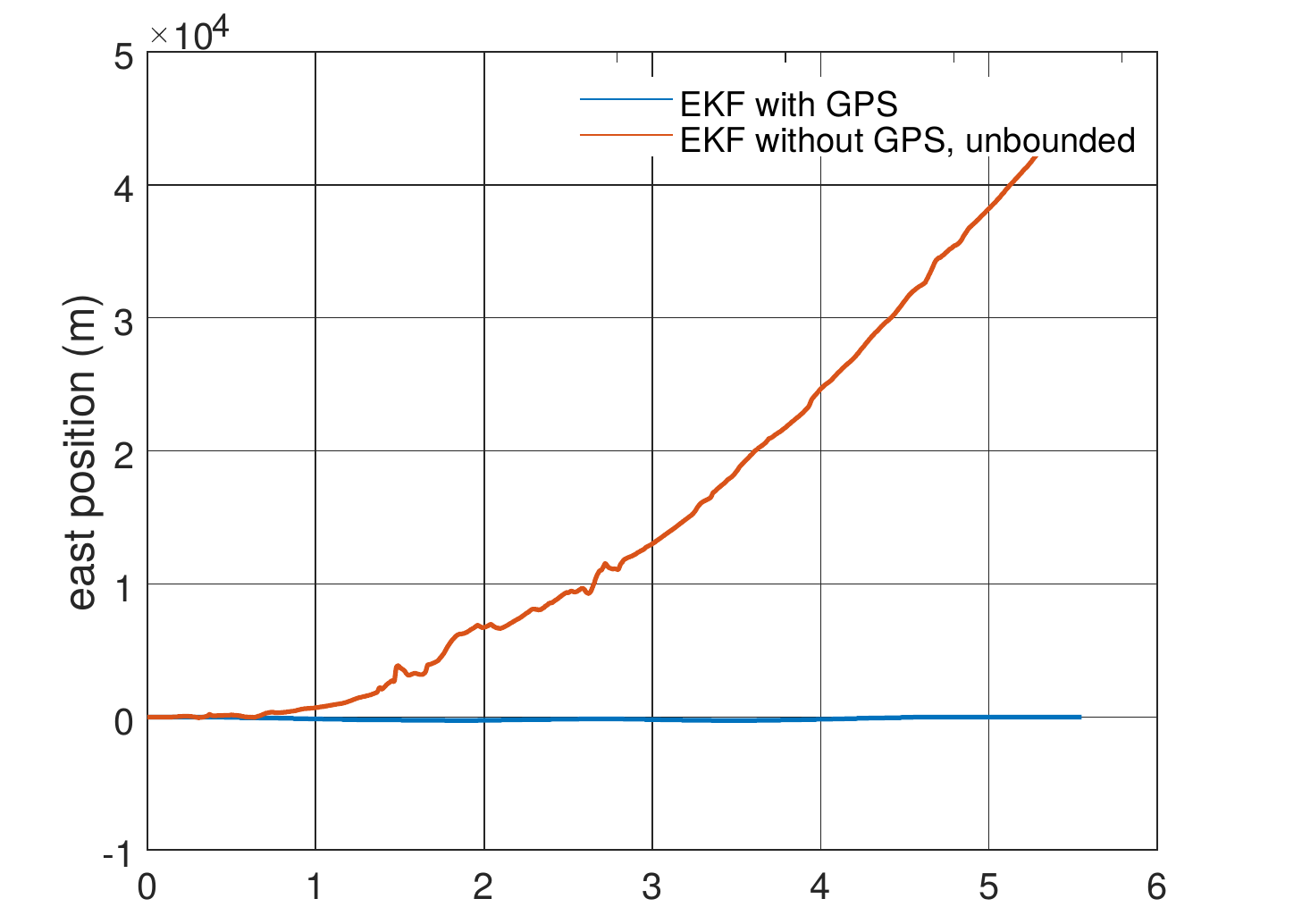}
	\caption{EKF2 east position predictions when no GPS aiding is used}
	\label{fig:no_gps_ekf}
\end{figure}

These predictions diverge quickly from the actual state values because numerical integration is applied thrice, and the biases
used for compensation are not perfectly accurate \cite{Jang2020}.
Biases even vary with time, so they need to be updated regularly. These
problems have more significant effects when low-cost sensors are used because
their errors are not consistent and hard to model \cite{Grewal2020},
which is the case with commercial autopilots like the Pixhawk. Figure \ref{fig:no_gps_ekf}
shows the predictions of the PX4 EKF2 in a real flight when GPS is
not used to correct the IMU predictions. 

\begin{figure}[h]
	\centering
	\includegraphics[width=1\linewidth]{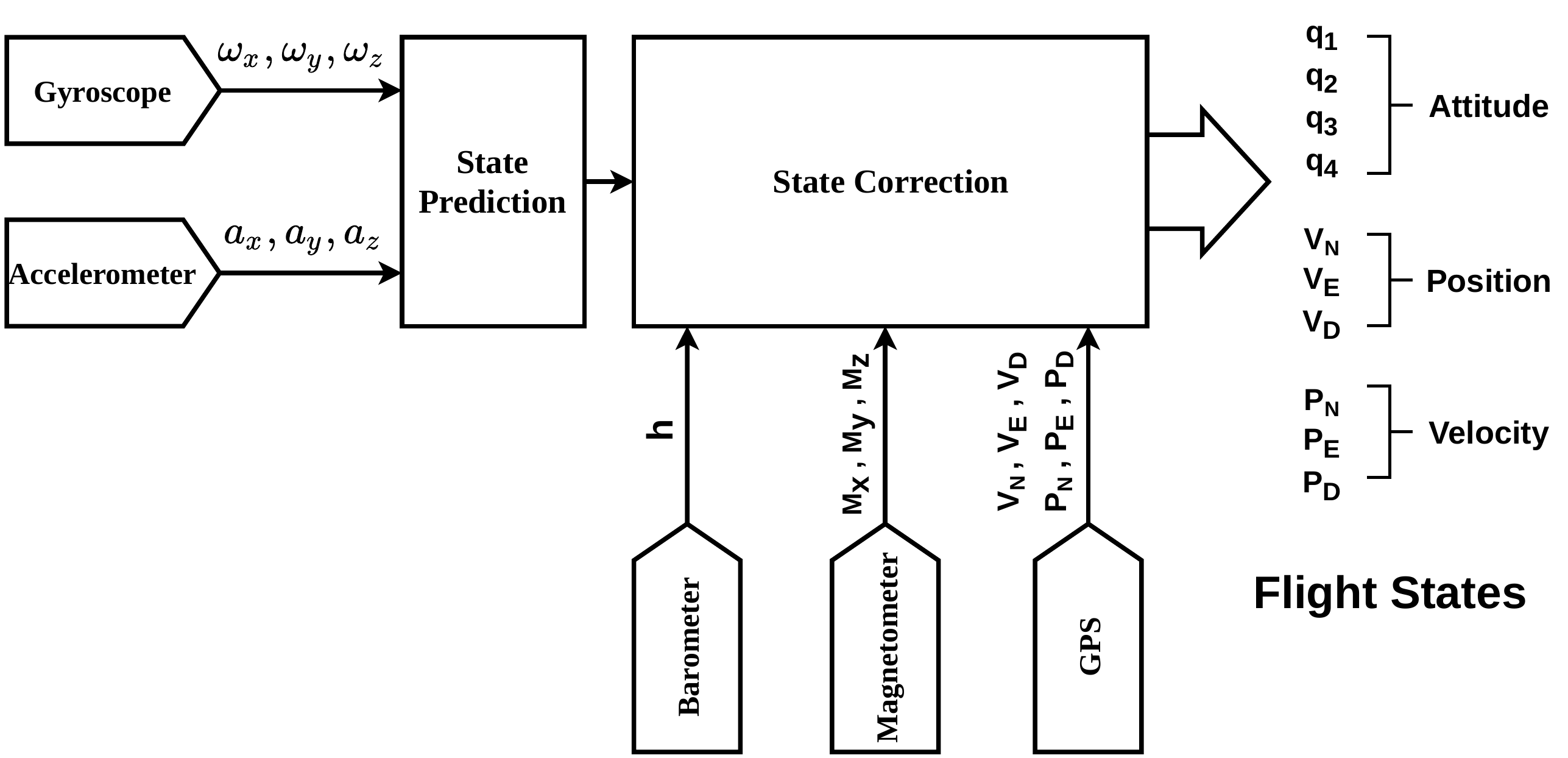}
	\caption{EKF Prediction-Correction Architecture}
	\label{fig:EKF_overview}
\end{figure}

The second step in EKF2 is the correction. Whenever measurements from
another sensor are available, the predictions from the first step
are corrected. The Barometer is used to correct the down component
of the position, while GPS corrects the other two components of position
and the three velocity components. The Magnetometer helps to correct
the attitude. Figure \ref{fig:EKF_overview} shows the
outline of the EKF prediction-correction architecture. 

Sensors other than the IMU are also used to correct the IMU biases
in the correction step to allow for better IMU predictions in the
next timestep. IMU bias correction is accomplished by adding the biases in delta
angles and delta velocities to the EKF state vector.

\section{Proposed System}

\begin{figure}[h]
	\centering
	\includegraphics[width=1\linewidth]{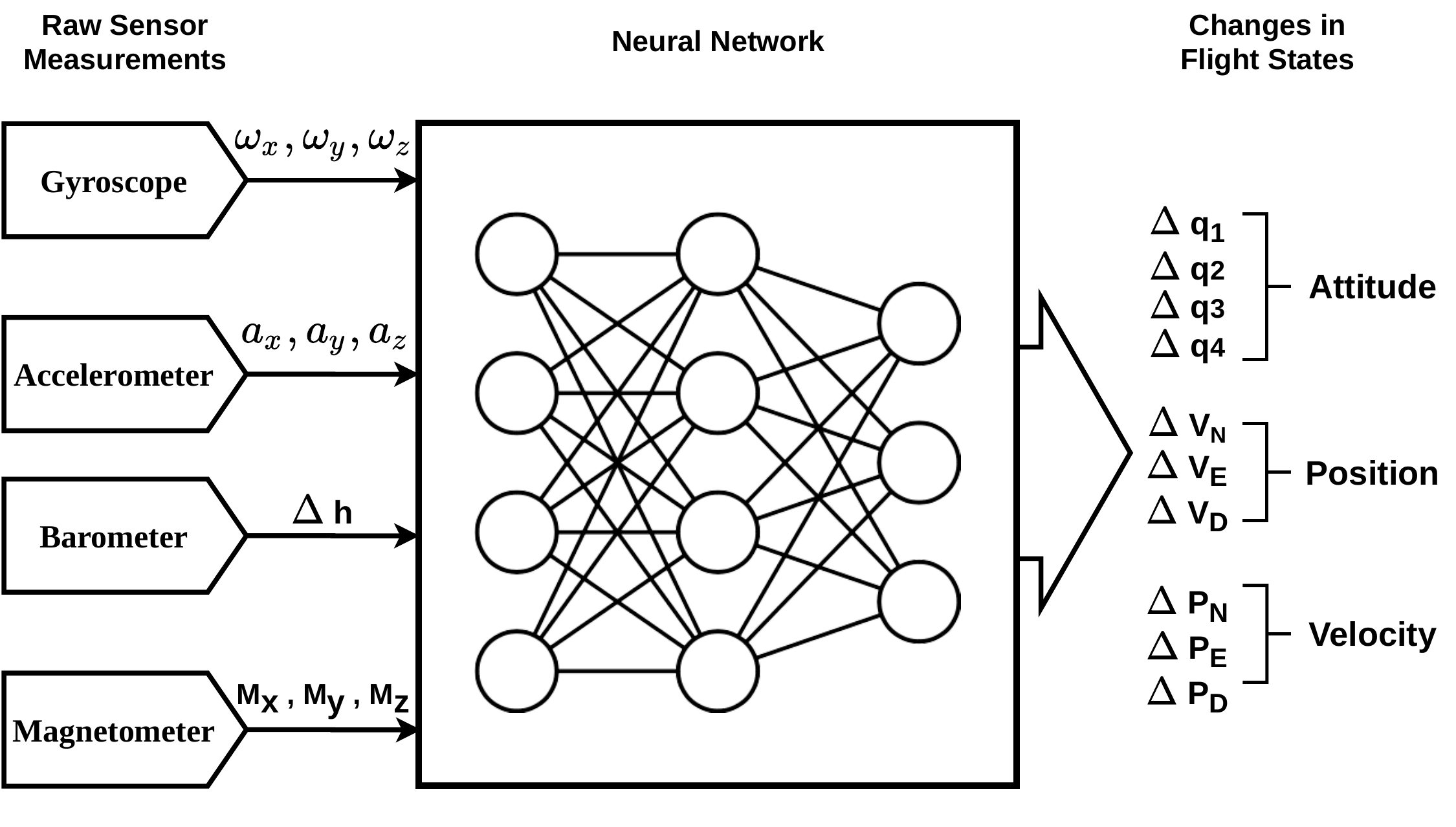}
	\caption{Proposed NN-Based Estimator }
	\label{fig:nn_estimation_overview}
\end{figure}

This paper proposes a new method to predict the states from the IMU
measurements without a need for GPS correction, sensor calibration,
or navigation equations. The main idea is to replace the processes shown
in Figure \ref{fig:inertial_nav_gyro} and Figure \ref{fig:inertial_nav_accel}
with a Neural Network NN, Figure \ref{fig:nn_estimation_overview}. The network is
basically learning the sensors' error models, measurement rotation,
and integration. 

The NN in this context can be viewed as a complex curve fitting function.
In curve fitting, a function, polynomial for example, is established
to represent the relation between an independent variable, like time,
and a dependent variable, like the price of an item. The coefficients
of this polynomial are calculated from a collection of input-output
pairs, that is, a table of time and price points (training set). Once
the polynomial coefficients are calculated, new values of time can
be substituted into the polynomial to predict the price, for example,
in the future. New inputs with known outputs are then given to the computed
fitting function to test its validity.  This test examines the accuracy on 
data not seen previously during fitting (validation set).
NNs follow the same concept but can fit more
complex relations using combinations of nonlinear functions and a large
number of parameters, called weights. In this case, the independent
variables (NN inputs) are the IMU, Barometer, and Magnetometer measurements.
The dependent variables (NN outputs) are the changes in position and
velocity.

The network architecture shown in Figure \ref{fig:nn_estimation_overview} is called
dense architecture. A better alternative when inputs or outputs are
in data series form is the Recurrent Neural Network (RNN).
RNNs are better suited for series of arbitrary lengths and are popular
with speech recognition and time series forecasting applications.
RNNs pass memory information from earlier points in the sequence to
later points and use them for later predictions. The architecture
used in this paper is composed of recurrent layers followed by a single
dense layer.

\section{Data Preprocessing}

No filtration is applied to the raw sensor measurements given to the
network. All the preprocessing steps mentioned here are used to control
the input rates, remove corrupted data from the training set, or divide
the inputs into segments.

\subsection{Different Update Rates of Different logged messages}

\begin{figure}[h]
	\centering
	\includegraphics[width=1\linewidth]{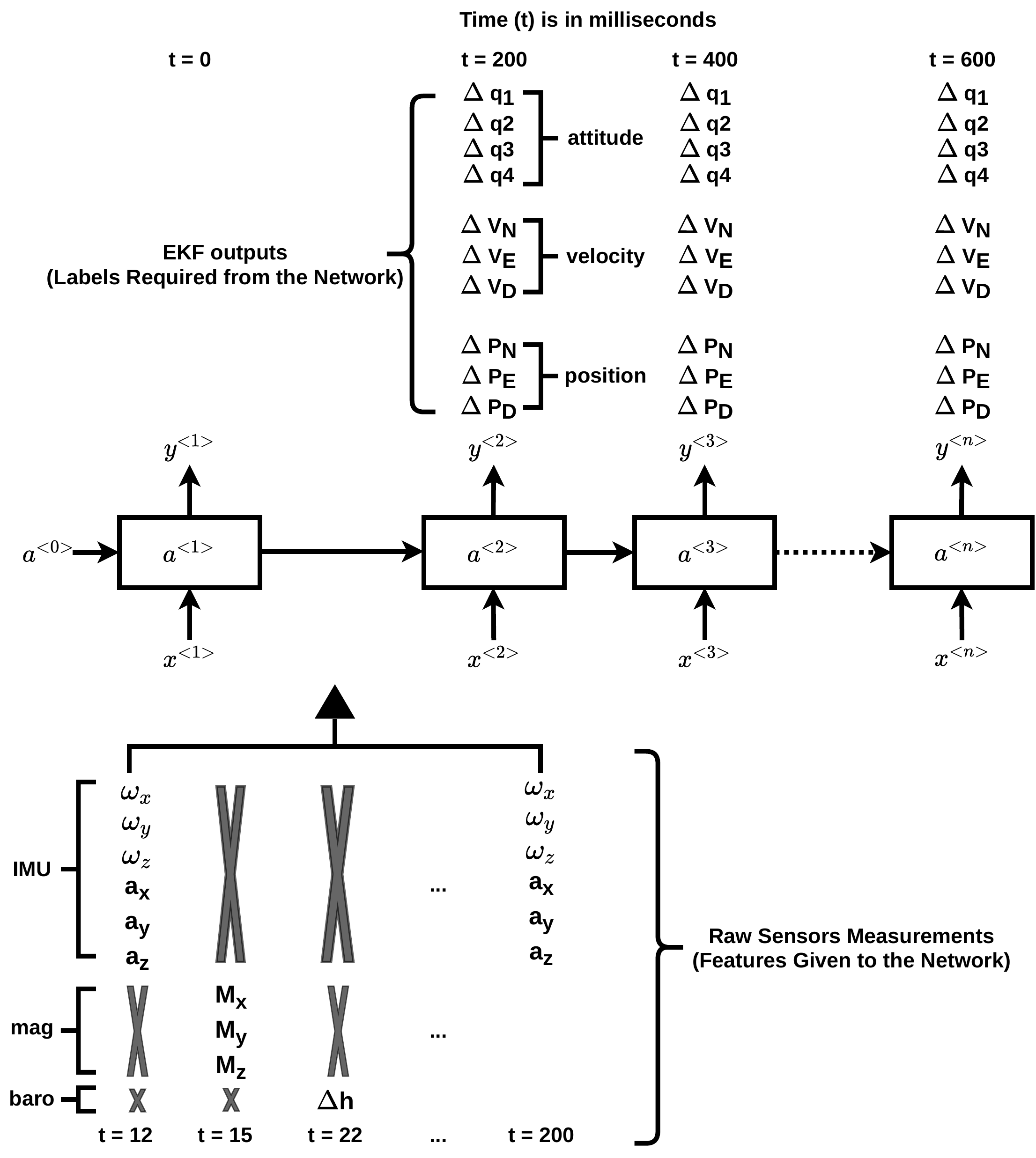}
	\caption{Unifying the different rates of features and labels retrieved from the flight logs}
	\label{fig:nn_multiple_rates}
\end{figure}

In the flight log file created by the PX4 stack, different sensor
data are stored in different messages. The frequency of the logged
data is not the same in all messages; different frequencies are shown
in Table \ref{tab:different_sensors_rates}. This means that between two
consecutive outputs expected from the network, several inputs arrive.
Furthermore, the input size differs from one step to another; for
example, only the IMU measurements are available at one step, but
in the next step, only Barometer data might be available, and so on.
To solve this problem, sensor data are averaged between each two consecutive
logged EKF outputs (Figure \ref{fig:nn_multiple_rates}). This averaging unifies
the input and output frequency to 5 Hz.

\begin{table}[h]
	\caption{Different sensors rates in the log file }
	\centering
	\setlength{\tabcolsep}{1pt}
	\begin{tabular}{| M{0.25\linewidth} | M{0.55\linewidth} | M{0.15\linewidth} |}
		\hline 
		Message & Contents & Update Rate \\
		\hline 
		\hline 
		IMU & Angular Rates $\omega_{x},\omega_{y},\omega_{z}$and Linear Accelerations
		$a_{x},a_{y},a_{z}$in body frame & 84 Hz \\
		\hline 
		Barometer & Temperature $T$ and Altitude $h$ & 67 Hz\\
		\hline 
		Magnetometer & Magnetic Field Components $M_{x},M_{y},M_{z}$in body frame & 45 Hz\\
		\hline 
		EKF & Estimated states, attitude $q_{1},q_{2},q_{3},q_{4}$, Velocity $V_{N},V_{E},V_{D}$
		and Position $P_{N},P_{E},P_{D}$ in Local Frame & 5 Hz\\
		\hline 
	\end{tabular}
	\label{tab:different_sensors_rates}
\end{table}

\subsection{Data Windowing}

RNNs are used with sequences because they learn the dependencies in
sequences of data. The length of the sequence fed into the network
at a time is called window size. For example, if the network 
translates a text from one language to another, the window might be
one sentence, paragraph, or page. In this problem, the window
length is defined by the number of timesteps, knowing that each timestep
is 0.2 seconds. Optimal window size depends on the problem and is
considered a design parameter and usually selected by trials. Large
windows require a longer time for both training and inference, and smaller
windows might not capture enough dependencies.

\subsection{Differencing}

Inertial sensors measure rates, so they can only be used to calculate
increments in flight states (position and velocity). Then the absolute
values of the states are calculated by accumulating these increments
to the initial conditions. This is why it is better to have the network
predict increments and accumulate the predictions later. So, in preprocessing,
differencing is applied to the labels (position, velocity, and attitude)
and to the input Barometer altitude. Differencing is done by subtracting
every point in the sequence from the next one.

\begin{figure}[h]
	\centering
	\includegraphics[width=1\linewidth]{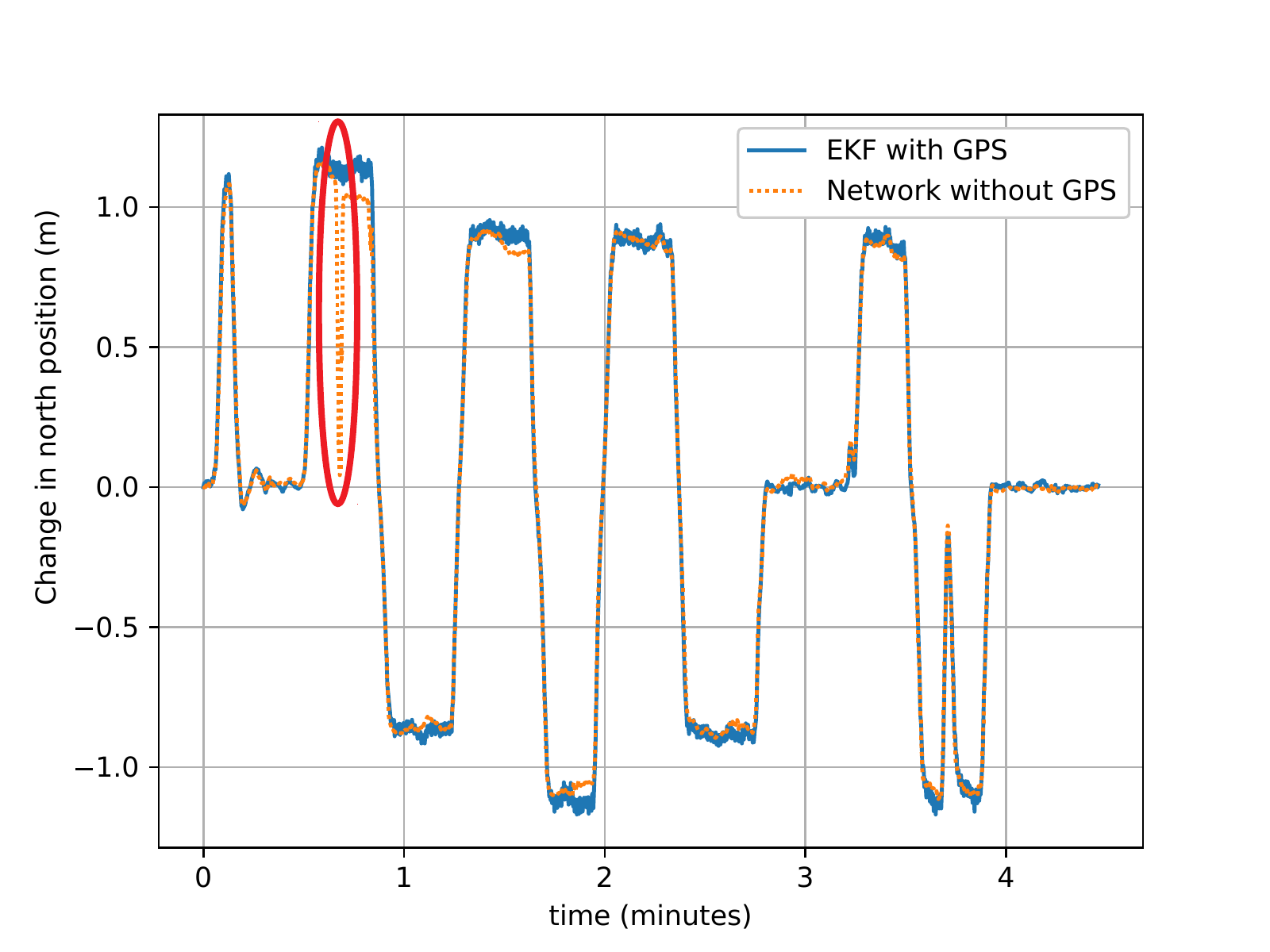}
	\caption{Changes in a state as predicted by the network as opposed to the true changes}
	\label{fig:local_error_diff}
\end{figure}

\begin{figure}[h]
	\centering
	\includegraphics[width=1\linewidth]{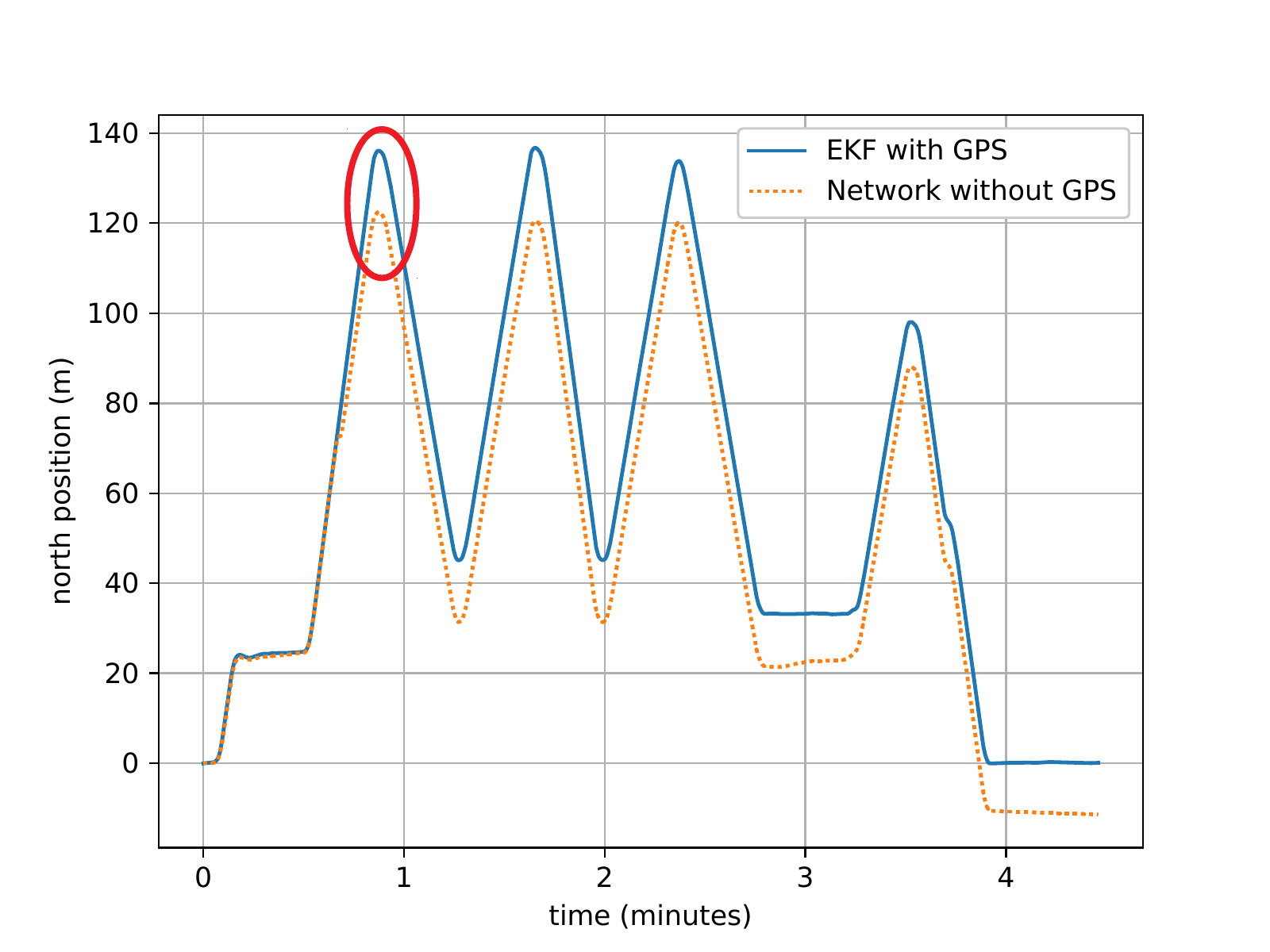}
	\caption{A mispredicted change results in an offset when reconstructing the state path, but it does not grow}
	\label{fig:local_error_recon}
\end{figure}

Another advantage of this approach is that a prediction at a time
step is not affected by an erroneous prediction of a previous step;
only the reconstructed path is affected. As a result, the errors in
the predicted states will be local, and though they cause a constant
offset between the true and predicted states for the rest of the flight,
this error \textit{does not grow} with time. Figure \ref{fig:local_error_diff}
and Figure \ref{fig:local_error_recon} show how an erroneous prediction
results in a bounded error that does not grow with time. This is only
possible because changes in states are being predicted, not the states
themselves. The offset grows up when multiple consecutive
errors in the same direction are made, and grows down when errors
of opposing direction are made.

\subsection{Ground Time Trimming} \label{subsec:ground_time_trimming}

\begin{figure}[h]
	\centering
	\includegraphics[width=1\linewidth]{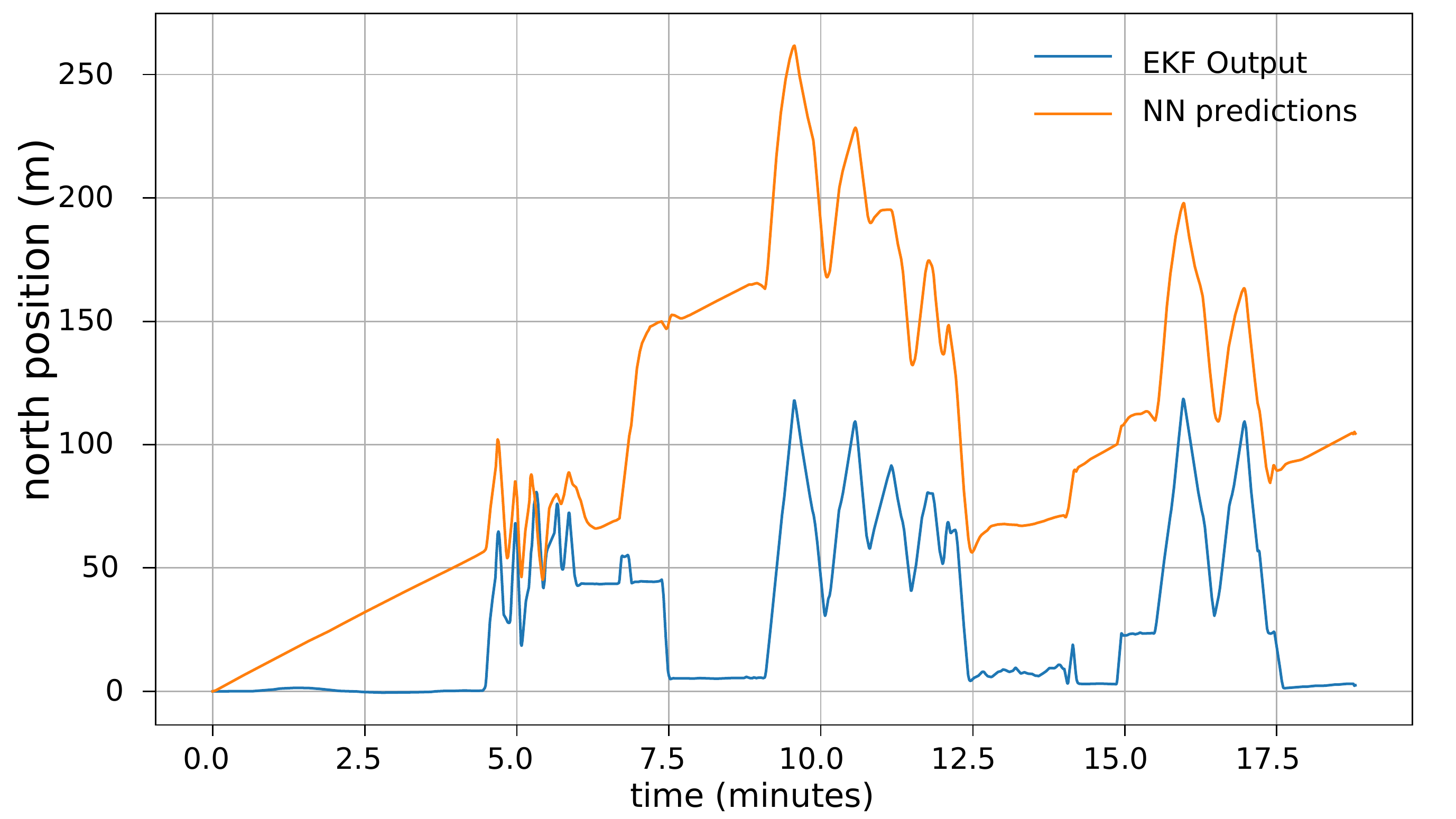}
	\caption{Network predictions drifting when the true values are not changing}
	\label{fig:drifting_predictions}
\end{figure}

While tuning the network, it was found that the network ``leaks'',
that is, when the true values are not changing, the network drifts
slowly. But when the true values are changing, the network predictions
follow the changes correctly. This can be seen in Figure \ref{fig:drifting_predictions}.
The network learned this behavior from the ground truth values. It turns
out that the EKF2 estimates drift when the drone is on the ground,
before takeoff and after landing, even if a good GPS signal is available. 

\begin{figure}[h]
	\centering
	\includegraphics[width=1\linewidth]{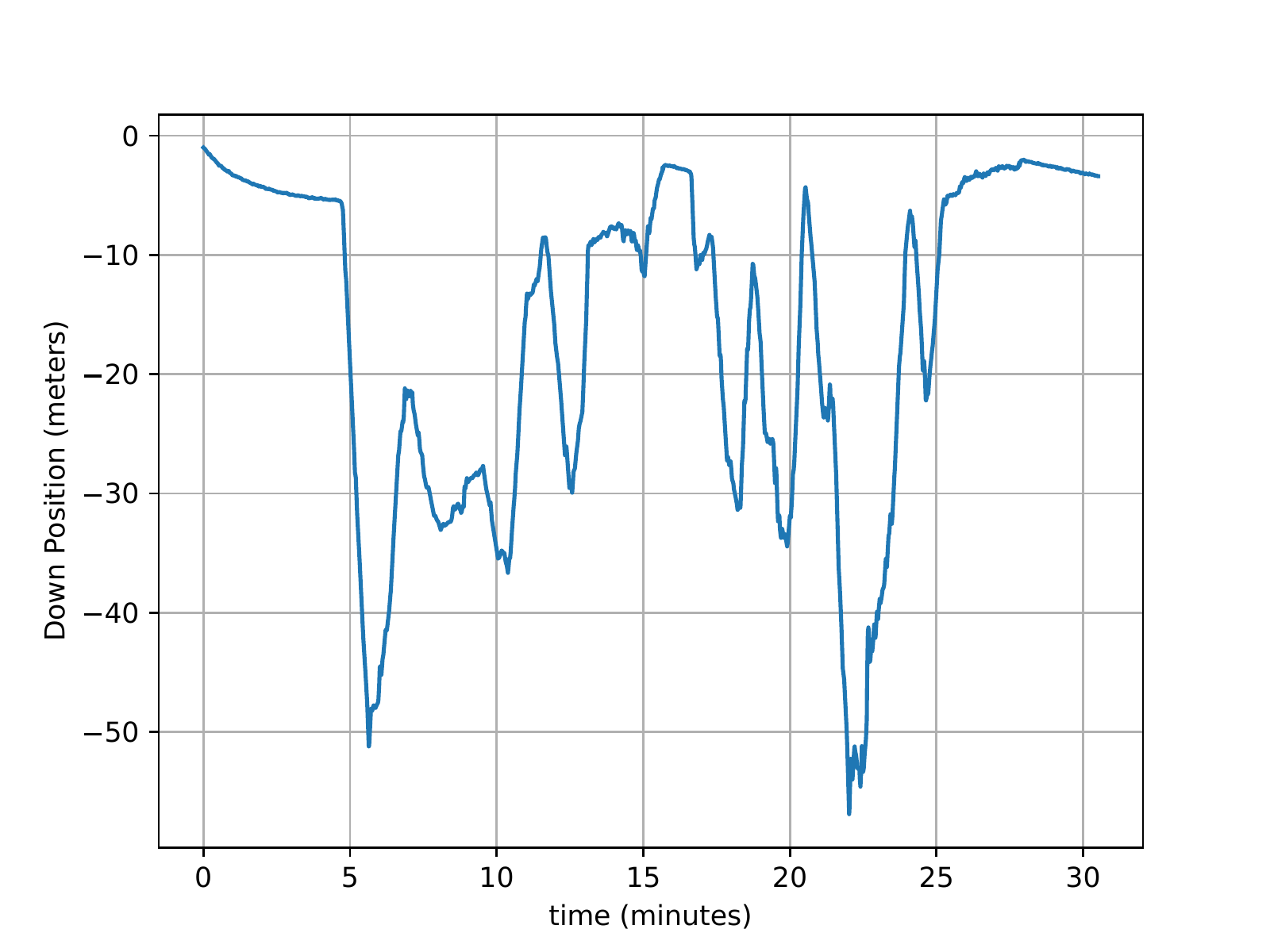}
	\caption{EKF estimates drifting when the drone is on the ground}
	\label{fig:drifting_ground}	
\end{figure}

Figure \ref{fig:drifting_ground} is an example of a thirty-minute
log of which about ten minutes are erroneous ground truth values.
This behavior exists in most flights, so it is difficult
for the network not to learn it. This behavior can be corrected by
trimming the ground time from the logs before feeding them to the
network.

\subsection{Manual Dataset Cleanup}

\begin{figure}[h]
	\centering
	\includegraphics[width=1\linewidth]{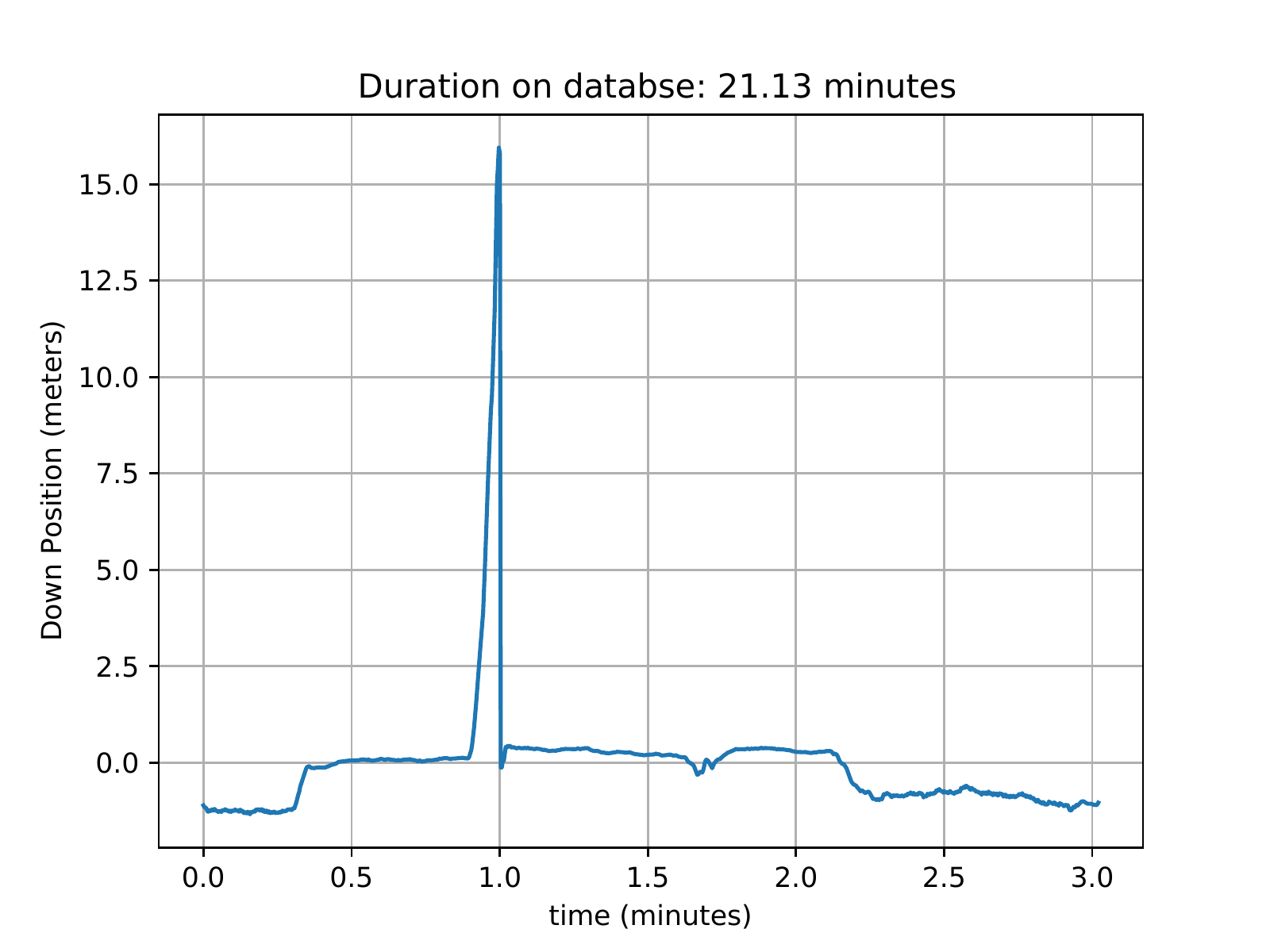}
	\caption{Corrupted log with no actual takeoff}
	\label{fig:corrupted_flight}
\end{figure}

Further manual cleanup was applied to remove other problems in the
dataset. For example, some logs where the drone did not even takeoff
might be as long as 20 minutes. Other logs collected with a poorly
tuned EKF, or when a hardware fault occurred are also removed. This
intensive cleanup reduced the number of logs from 943 to 548. Figure \ref{fig:corrupted_flight}
shows a log found in the database with a claimed duration of about
21 minutes. After trimming the ground time it was found that the entire
log is corrupted, and no actual takeoff took place.

\section{Network Design and Hyperparameters	Tuning} \label{sec:network_design}

\begin{figure}[h]
	\centering
	\includegraphics[width=1\linewidth]{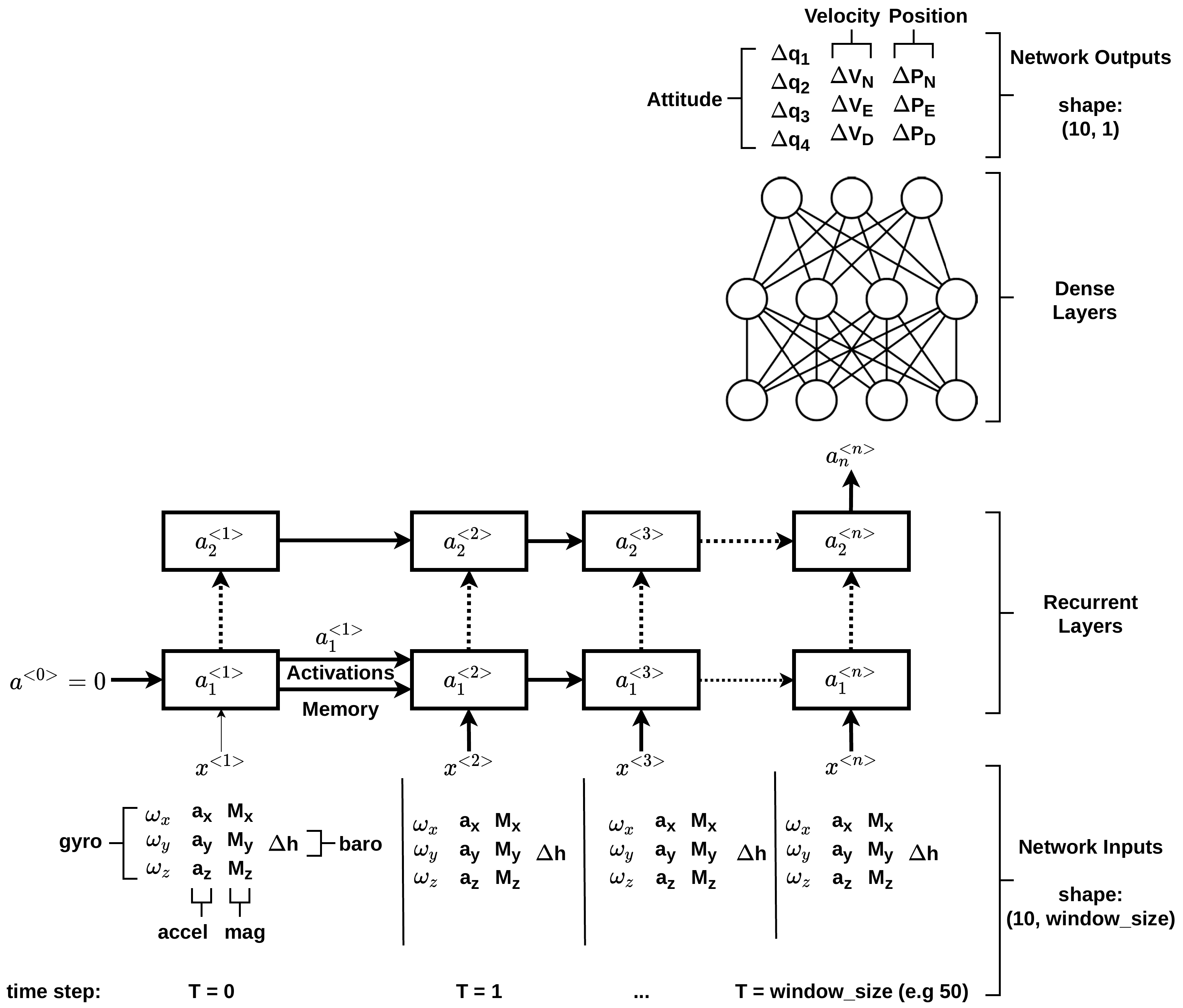}
	\caption{General Network Architecture}
	\label{fig:nn_general_outline}
\end{figure}

Neural Network design is an iterative process. Many design parameters
that determine the network shape and characteristics are to be selected.
Other design decisions are concerned with how the network is trained,
that is, how the weights are updated. These decisions result in the
selection of proper \textit{hyperparameters}. This section describes
how each design parameter and hyperparameter was selected. Figure \ref{fig:nn_general_outline}
shows the general network architecture. The number of layers, their
types, and the number of nodes in each layer are some of the design
decisions.

To decide on hyperparameters, a performance index is needed, but the
system can be assessed from two different perspectives
\begin{enumerate}
	\item As a neural network
	
	In this case, a good network makes as many correct predictions as
	possible. Recall that predictions are changes in the states, not the
	states themselves. Correct prediction means that the predicted value is
	close enough to the true value because exact correctness is not
	feasible with floats.
	
	\item As a Navigation System
	
	In this case, a good design results in the smallest difference between
	the predicted and the true paths. It can also be assessed by
	the growth of error with time. In this case, the reconstructed states
	plots are visualized, not the predictions themselves. 
\end{enumerate}

If a design makes a single mistake at the beginning of a given flight,
a constant offset is kept between the true and estimated path
for the entire flight duration. This design is a good NN but is a
lousy navigation system. Another design might make multiple mistakes
at the end of a flight, so their effect on the path will be negligible.
This design is better from a Navigation perspective but is a worse NN.

During training, only the first perspective can be considered through
the loss (cost) value. The second assessment method requires post-processing
after the network is trained. The 3D path of every validation
flight is compared to the true 3D path. The maximum difference for
each flight is recorded. The median of these maximum values is
calculated and called MPE. The selection of the upcoming hyperparameters
was made according to MPE as a single performance index.

\subsection{Loss Function}

\begin{figure}[h]
	\centering
	\includegraphics[width=1\linewidth]{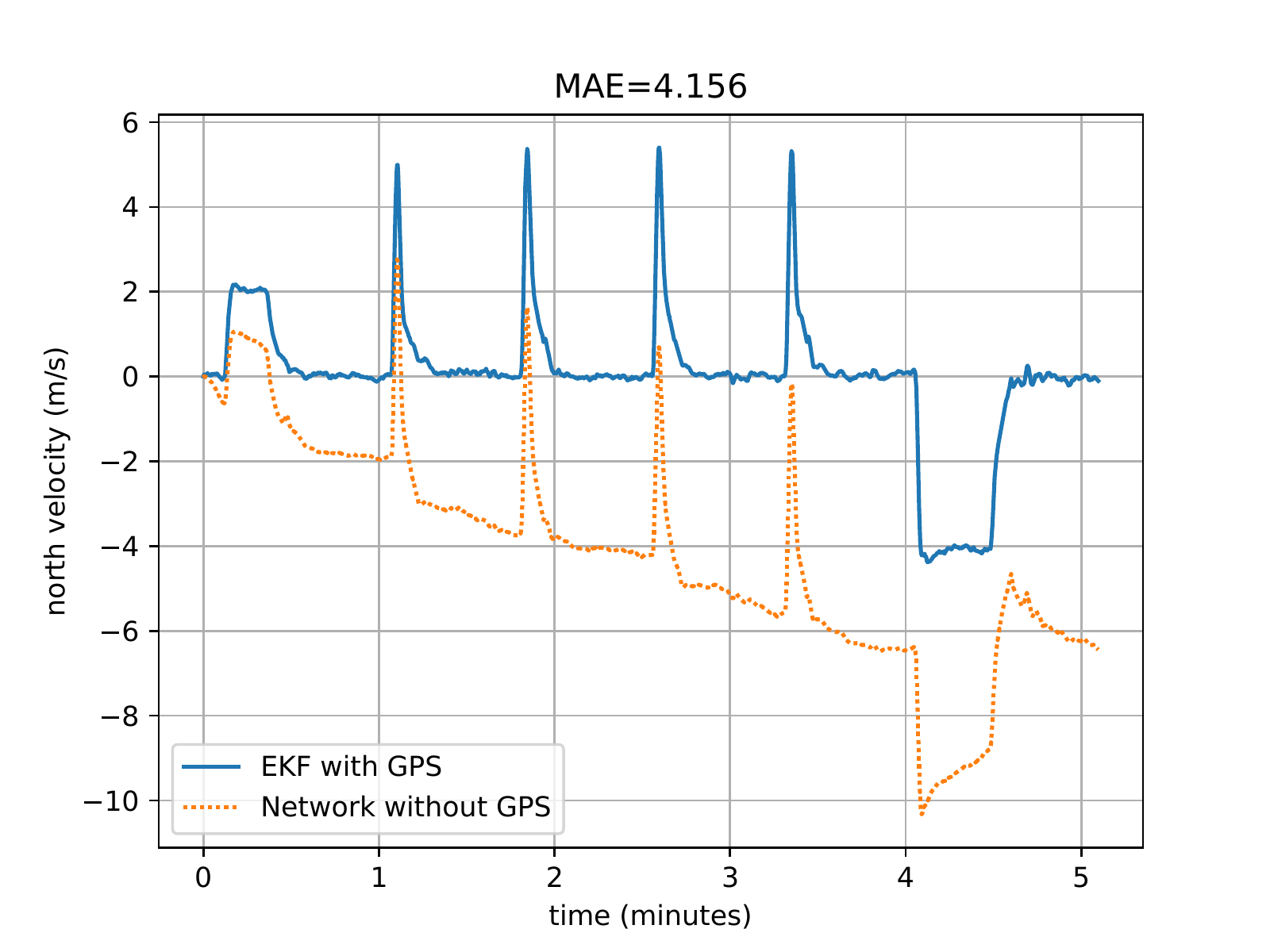}
	\caption{Less attention is paid to velocity errors due to the smaller magnitudes}
	\label{fig:bad_north_velocity}
\end{figure}

\begin{figure}[h]
	\centering
	\includegraphics[width=1\linewidth]{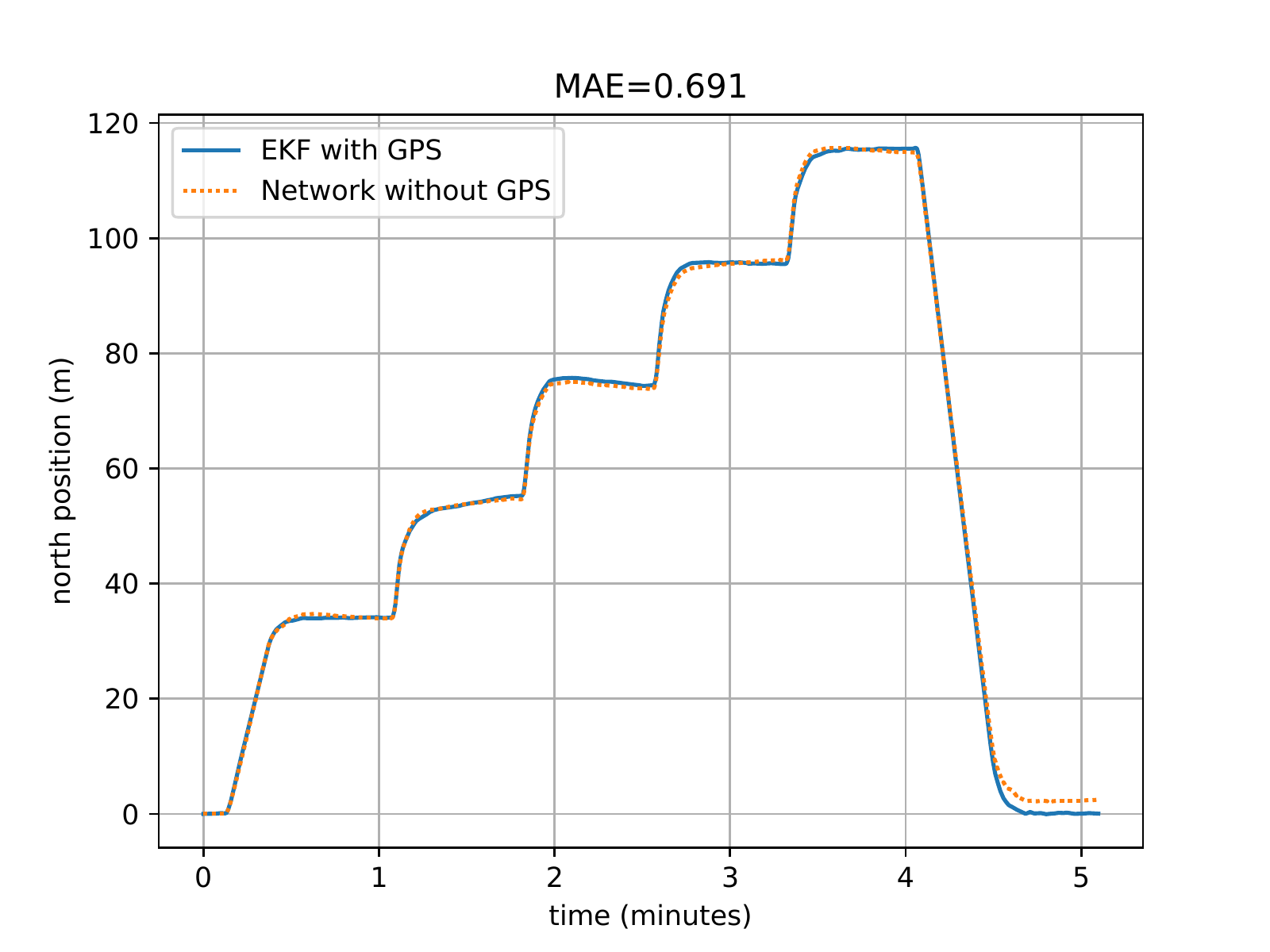}
	\caption{More attention is paid to the large position components, so higher accuracy is acheived}
	\label{fig:good_north_position}
\end{figure}

The loss function determines how far the prediction is from
the true value, thus, how much the weights need to be changed. Standard
loss functions used with timeseries forecasting like Mean Absolute
Error (MAE), Mean Square Error (MSE), and Huber Loss did not perform well in this particular problem. This is because the output
signals of the network have different orders of magnitude. For example,
velocity components are bounded by the vehicle's propulsive and structural
capabilities. They are considerably smaller than the position components
that can grow as large as the drone's datalink to the ground station
can support. This translates to large velocity errors, Figure \ref{fig:bad_north_velocity}, even
when the position errors are small, Figure \ref{fig:good_north_position}.
However, position estimation is more complex than velocity estimation and implicitly utilizes it. This is the result of the 
network paying more attention to the signals that produce larger error magnitudes, thus have larger contribution to the cost.

To solve this, a modified MAE was used to account for different signals
magnitudes. Equation \ref{MAE} shows the standard MAE, where $\hat{y_{i}}$
is the state value predicted by the network, $y_{i}$ is the true
state value, and $n$ is the number of predicted values, $n=6$ if
only velocity and position are required and $n=10$ if the attitude
is required too. Equation \ref{MAE_weighted} is the weighted version, for
each signal, the error is multiplied by the signal's weight so that
small errors coming from small signals contribute more fairly to the
final loss. The weight for each signal is the reciprocal of its mean
in the training set.

\begin{equation} \label{MAE}
	\mathrm{MAE}=\frac{\sum_{i=1}^{n}\left|\hat{y_{i}}-y_{i}\right|}{n}=\frac{\sum_{i=1}^{n}\left|e_{i}\right|}{n}\label{eq:MAE}
\end{equation}

\begin{equation} \label{MAE_weighted}
	\mathrm{MAE_{weighted}}=\frac{\sum_{i=1}^{n}\left|\hat{y_{i}}-y_{i}\right|{\color{red}w_{i}}}{n}=
	\frac{\sum_{i=1}^{n}\left|e_{i}\right|{\color{red}w_{i}}}{n}\label{eq:MAE_weighted}
\end{equation}

Using Mean Absolute Percentage Error (MAPE) loss cancels the need
for weighting. However, MAPE divides the error by the true value which
is often zero. Catching zero divisions and replacing them with machine
precision did not solve the problem because the resulting errors were
too high. MSE was not better than MAE because most values are small,
and squaring made them even smaller resulting in small loss values.

\subsection{Network Architecture}

Different combinations of dense and recurrent layers
were tested, from a single recurrent layer of 10 nodes to 12 layers
of 200 nodes each. The best performance was achieved by four recurrent
layers, 200 nodes each, followed by a single dense layer of 6 nodes
to reshape the data to the shape of the labels (if only velocity and
position are required).

\subsection{Recurrent Cell Type and Activation Functions}

Three different types of recurrent cells were tested; the Vanilla
Cell, the Gated Recurrent Unit (GRU), and the Long Short Term Memory
(LSTM) cell. Despite taking longer to train and resulting in more parameters, the LSTM cells gave the best performance. Two
activations are applied within an LSTM cell, one for the new inputs
and another for the recurrent inputs. Sigmoid, tanh and ReLU activations
were tested. While tanh had the best performance, it could only be
used for input activation. The recurrent activation must be sigmoid
for the GPU accelerated training to work.

\subsection{Window Size}

Window sizes from 20 steps to 200 steps were tested. Increasing the
window size increases the accuracy but significantly increases both training
and inference time. Too large windows might decrease the accuracy
because a larger than needed context leads to network confusion. Table \ref{tab:window_size_effect} 
shows the effect of inreasing window size on both accuracy and training time.

\begin{table}[h]
	\label{tab:window_size_effect}
	\caption{Effect of increasing window size on accuracy and training time}	
	\centering
	\setlength{\tabcolsep}{1pt}
	\begin{tabular}{| M{0.15\linewidth} | M{0.45\linewidth} |  M{0.35\linewidth}|}
			\hline 
			Window size & Median of maximum position errors in validation set (meters) & Training time, 100 epochs (hours)\\
			\hline 
			\hline 
			50 & 72.4 & 1.94\\
			\hline 
			100 & 52.78 & 4.94\\
			\hline 
			150 & 35.59 & 7.25\\
			\hline 
			200 & 34.26 & 9.21\\
			\hline 
	\end{tabular}

	\end{table}

\subsection{Learning Rate}

Learning rates from 0.00001 to 0.1 were tested. In the initial epochs,
when the weights are far from their optimal values, high learning
rate can be used. When the loss approaches its minimum, the learning
rate should be decreased to prevent overshooting. Table \ref{tab:learning_rate_schedule}
was used as a learning rate schedule.

\begin{table}[h]
	\caption{Learning rate schedule}
	\centering
	\begin{tabular}{|c|c|}
		\hline 
		Epochs (from - to) & Learning rate\\
		\hline 
		\hline 
		0 - 50  & 0.005\\
		\hline 
		51 - 100  & 0.0025\\
		\hline 
		101 - ...  & 0.001\\
		\hline 
	\end{tabular}
	\label{tab:learning_rate_schedule}
\end{table}

\subsection{Batch Size}

The number of training examples (windows) fed to the network in one
optimizer step is called batch size. Batching was initially used because
GPU memory could not accommodate the entire dataset. Using batches
can lead to faster convergence because of the more frequent weights
updates. Batch sizes from 512 to 16,384 windows were tested. The chosen
batch size is 1024, as it achieves the fastest learning and provides
some regularization.

\subsection{Regularization}

Regularization is needed to prevent the network from memorizing the training data and allow it to generalize to the validation data.
The most popular regularization technique is dropout, which randomly
turns off hidden nodes in training time when propagating each input
vector. Since two activations are applied, two dropouts are applied,
one for the input and another for the recurrent activations. In the
earlier trials, dropout gave very good regularization when small number of flight logs was used. Unfortunately, recurrent dropout prevents
GPU acceleration, so, it could not be used with large datasets. L1 and
L2 regularization were also tested, but they could not achieve similar
performance as dropout. The final design and the results shown in
this paper do not use any formal regularization.

\subsection{Training Time and Number of epochs}

An epoch is a single pass through the entire dataset, that is, processing
all the batches. Longer training increases training accuracy but may
reduce validation accuracy when overfitting occurs. Different numbers
of epochs were tested from 100 to 1500, but the accuracy did not increase
much after 200 epochs. This takes about 5 hours using two Nvidia RTX
2070 Super GPUs connected via NVLink.

\subsection{Final Design}

The final network design and hyperparameters used are summarized in
Table \ref{tab:network_architecture_and_hyperparam}.

\begin{table}[h]
	\caption{Network Architecture and Hyperparameters}
	\centering
	\setlength{\tabcolsep}{1pt}
	\begin{tabular}{| M{0.7\linewidth} | M{0.3\linewidth} |}
		\hline 
		Number of recurrent layers & 4\\
		\hline 
		Number of nodes in a recurrent cell & 200\\
		\hline 
		Input activation in recurrent layers & tanh\\
		\hline 
		Recurrent activation in recurrent layers & sigmoid\\
		\hline 
		Number of dense layers & 1\\
		\hline 
		Number of nodes in the dense layer & 6\\
		\hline 
		Dense activation & None\\
		\hline 
		Optimizer & Adam\\
		\hline 
		Adam's first \& second moment decay rates & 0.9, 0.999\\
		\hline 
		Number of Epochs & 100 (\ensuremath{\approx} 5 hours)\\
		\hline 
		Learning rate schedule & 0.005, 0.0025, 0.001\\
		\hline 
		Regularization & None\\
		\hline 
		Window size & 200 steps = 40 seconds\\
		\hline 
		Batch size & 1024 examples\\
		\hline 
		Loss function & Weighted MAE (custom)\\
		\hline 
	\end{tabular}
	\label{tab:network_architecture_and_hyperparam}
\end{table}

\section{Results and Discussion}

The MPE index defined in section \ref{sec:network_design} is enough to
compare different designs given that the flights in both training
and validation sets are unchanged. But when comparing the accuracy
of a design in two different flights, MPE alone is not sufficient,
because different flights vary in duration and traveled distance.
Traditionally, an INS is assessed by the growth of positioning error
with time, because the major estimation errors in INS accumulate proportionally
to the time since positioning starts. A similar index is used in the
following discussions, obtained by dividing the MPE by the flight
duration to arrive at Time-Normalized Maximum Position Error (TN-MPE),
expressed in meters per minute (m/min).

\subsection{Performance on the Training Set}

Training performance is usually examined to make sure that the network
is at least fitting the training data. No enhancements should be applied
to the validation performance until the network sufficiently fits
the training data. However, training performance is not a sufficient
indicator, and the network must not be expected to provide similar
performance on newer data. Training performance might also be treated
as a goal to look for in the validation set when tuning the regularization
parameters, because it shows the potential of a particular design. Table \ref{tab:training_performance}
summarizes the performance on the training set, which consists of
465 flights. The shown median position error indicates that positioning
error never grew beyond 7.23 meters in 232 different flights.

\begin{table}[h]
	\caption{Performance on the training dataset}
	\centering
	\setlength{\tabcolsep}{2pt}
	\begin{tabular}{|c|c|c|c|}
		\hline 
		& MPE (m) & TN-MPE (m/min) & MVE (m/s)\\
		\hline 
		\hline 
		Mean & 16.78 & 2.98 & 3.78\\
		\hline 
		Median & 7.23 & 1.39 & 2.26\\
		\hline 
		Best Flight & 1.13 & 0.45 & 0.57\\
		\hline 
		Worst Flight & 1666.31 & 262.69 & 168.02\\
		\hline 
	\end{tabular}
	\label{tab:training_performance}
\end{table}

\subsection{Performance on the Validation Set}

Validation performance reflects how the system performs with new flights
not used during training. The validation dataset is comprised of 83
flights totaling about 8 hours. Table \ref{tab:validation_performance} summarizes
the performance on the validation set. Mean and median are calculated
from the maximum errors reached in every flight. It is noted that
the validation mean MPE and TN-MPE are five times larger than those
of the training. This resulted from the lack of proper regularization.

\begin{table}[h]
	\caption{Performance on the validation dataset}
	\centering
	\setlength{\tabcolsep}{2pt}
	\begin{tabular}{|c|c|c|c|}
		\hline 
		& MPE (m) & TN-MPE (m/min) & MVE (m/s)\\
		\hline 
		\hline 
		Mean & 85.79 & 17.79 & 6.40\\
		\hline 
		Median & 34.26 & 6 & 4.15\\
		\hline 
		Best Flight & 2.66 & 0.61 & 0.76\\
		\hline 
		Worst Flight & 871.54 & 142.08 & 67.65\\
		\hline 
	\end{tabular}
	\label{tab:validation_performance}
\end{table}

\begin{figure}[h]
	\centering
	\includegraphics[width=1\linewidth]{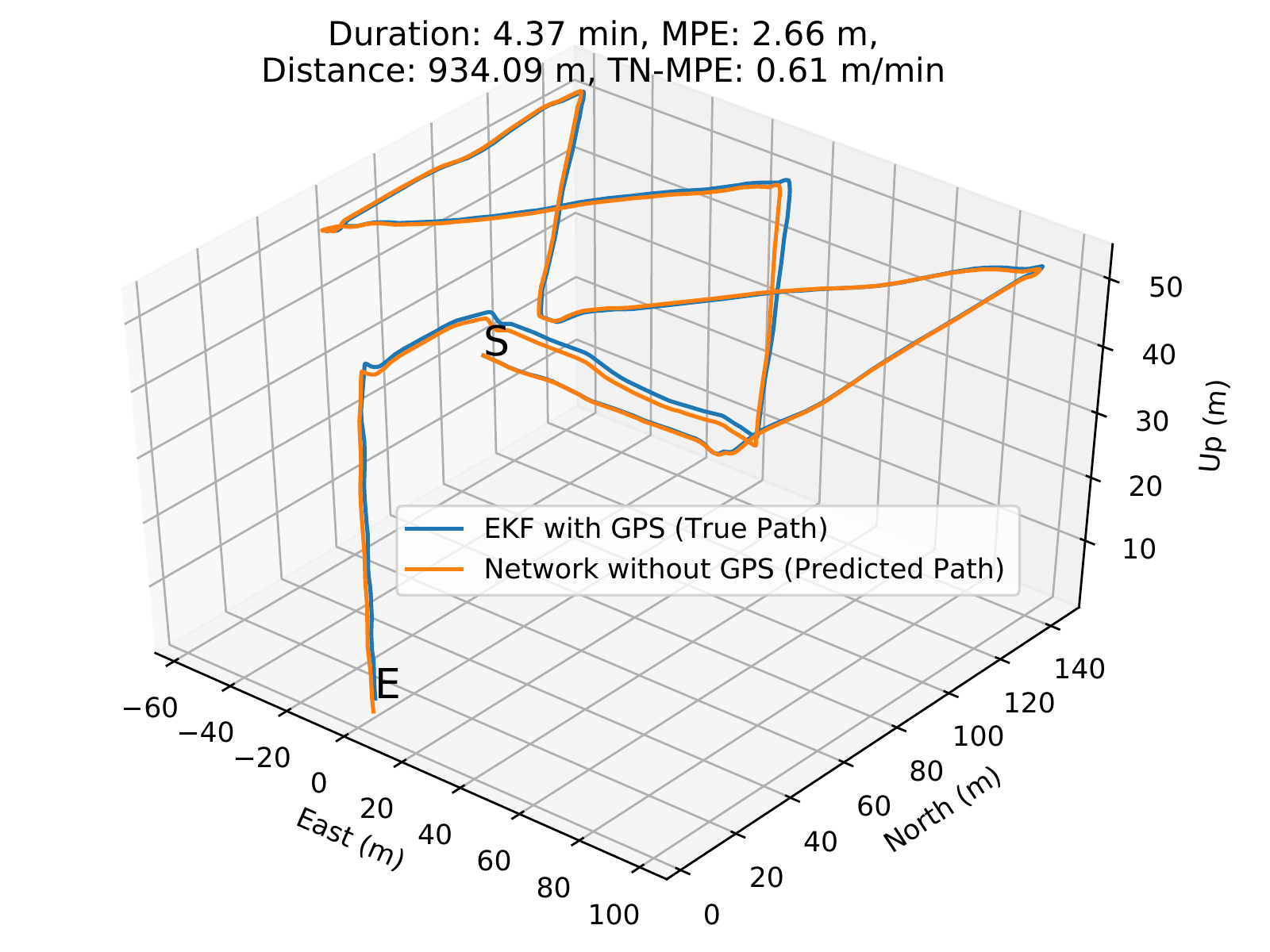}
	\caption{Path of the best validation flight}
	\label{fig:val_flight_triangles}
\end{figure}

Figure \ref{fig:val_flight_triangles} shows the flight path of the validation flight
with the lowest MPE and TN-MPE. The S letter in a plot indicates the
true start position and the E letter indicates the true end position.
Some flights seem to start mid-air because the ground time is trimmed
as described in section \ref{subsec:ground_time_trimming}, and the takeoff
portion lies within the first window. As mentioned in Table \ref{tab:network_architecture_and_hyperparam},
the window size is 200 steps. This means that the network makes its
first prediction after 40 seconds. By that time, the drone might have
already reached the mission altitude.

Despite the path in Figure \ref{fig:val_flight_triangles} having some acute direction
changes, the network could still predict these maneuvers correctly.
One reason is that this flight pattern is common in the dataset,
so the network had enough data to relate the associated sensor behavior
to the resulting changes in position. 

\begin{figure}[h]
	\centering
	\includegraphics[width=1\linewidth]{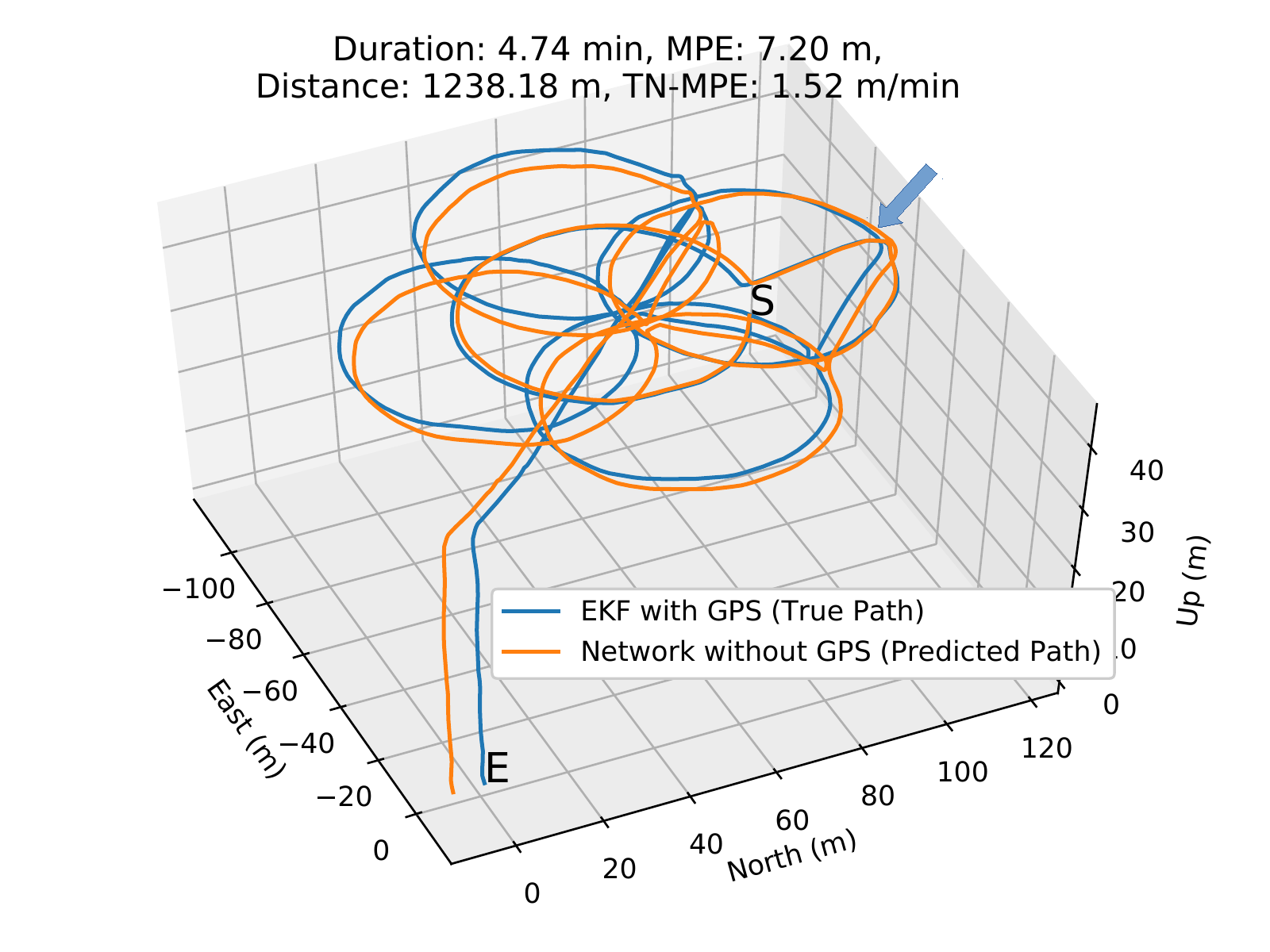}
	\caption{Validation flight path -1}
	\label{fig:val_flight_circles}
\end{figure}

Figure \ref{fig:val_flight_circles} shows another validation flight with
different maneuvers. The blue arrow marks a sudden change in turn
rate; an erroneous prediction is made at this point resulting in
an offset that persists until the end of the flight. 

\begin{figure}[h]
	\centering
	\includegraphics[width=1\linewidth]{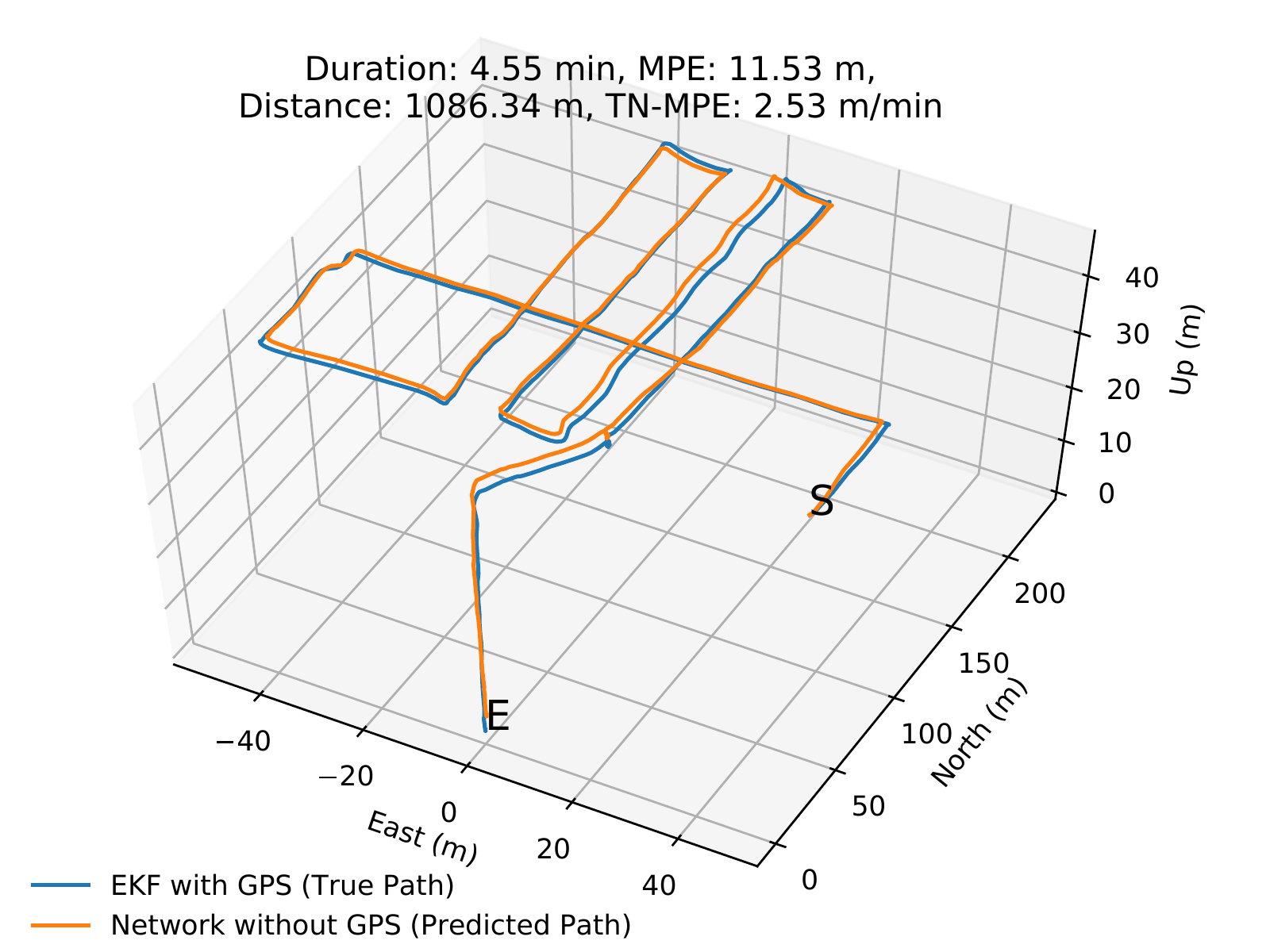}
	\caption{Validation flight path -2}
	\label{fig:val_flight_semi_survey}
\end{figure}

\begin{figure}[h]
	\centering
	\includegraphics[width=1\linewidth]{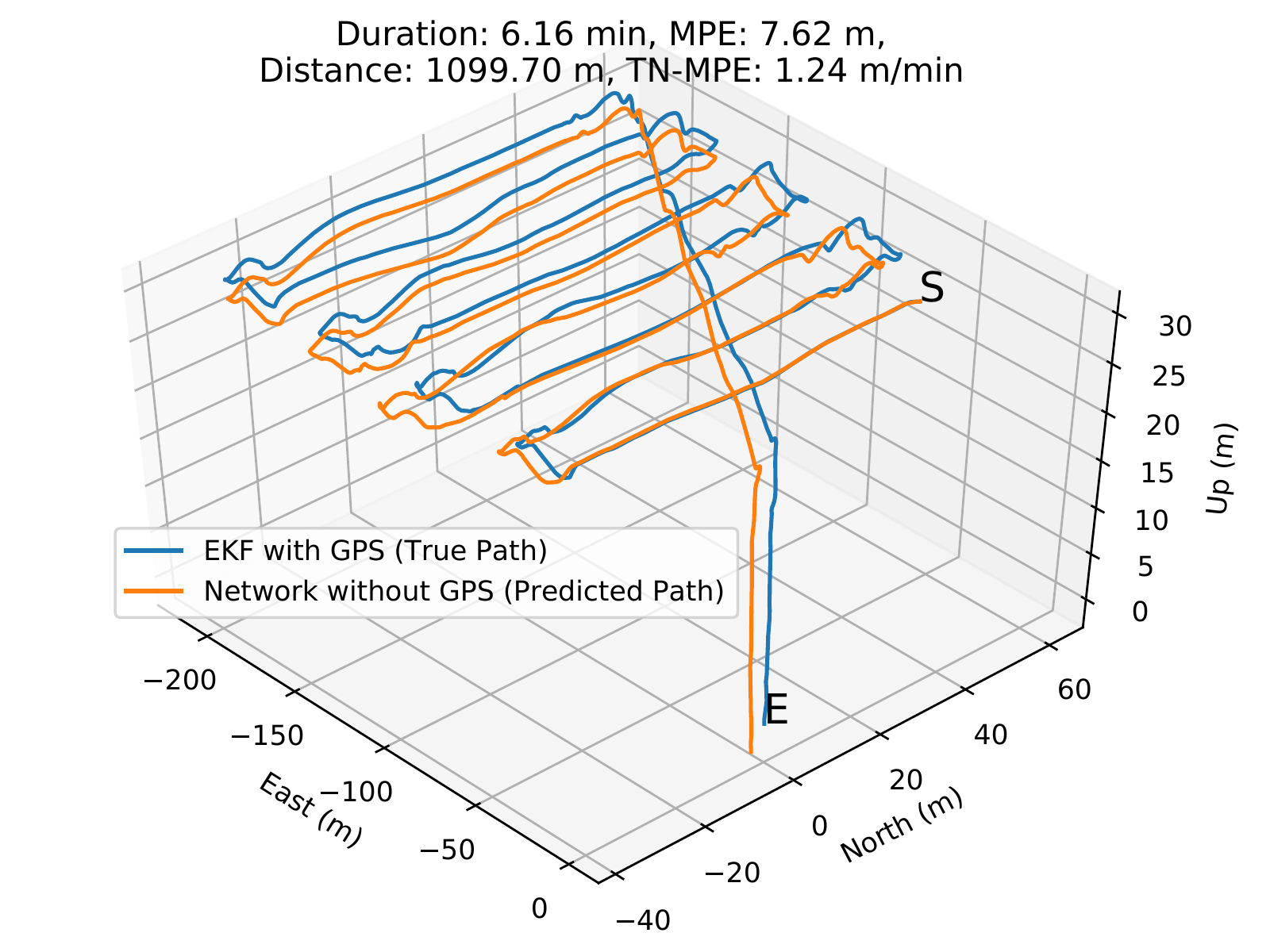}
	\caption{Validation flight path -3}
	\label{fig:val_flight_survey}
\end{figure}

\begin{figure}[h]
	\centering
	\includegraphics[width=1\linewidth]{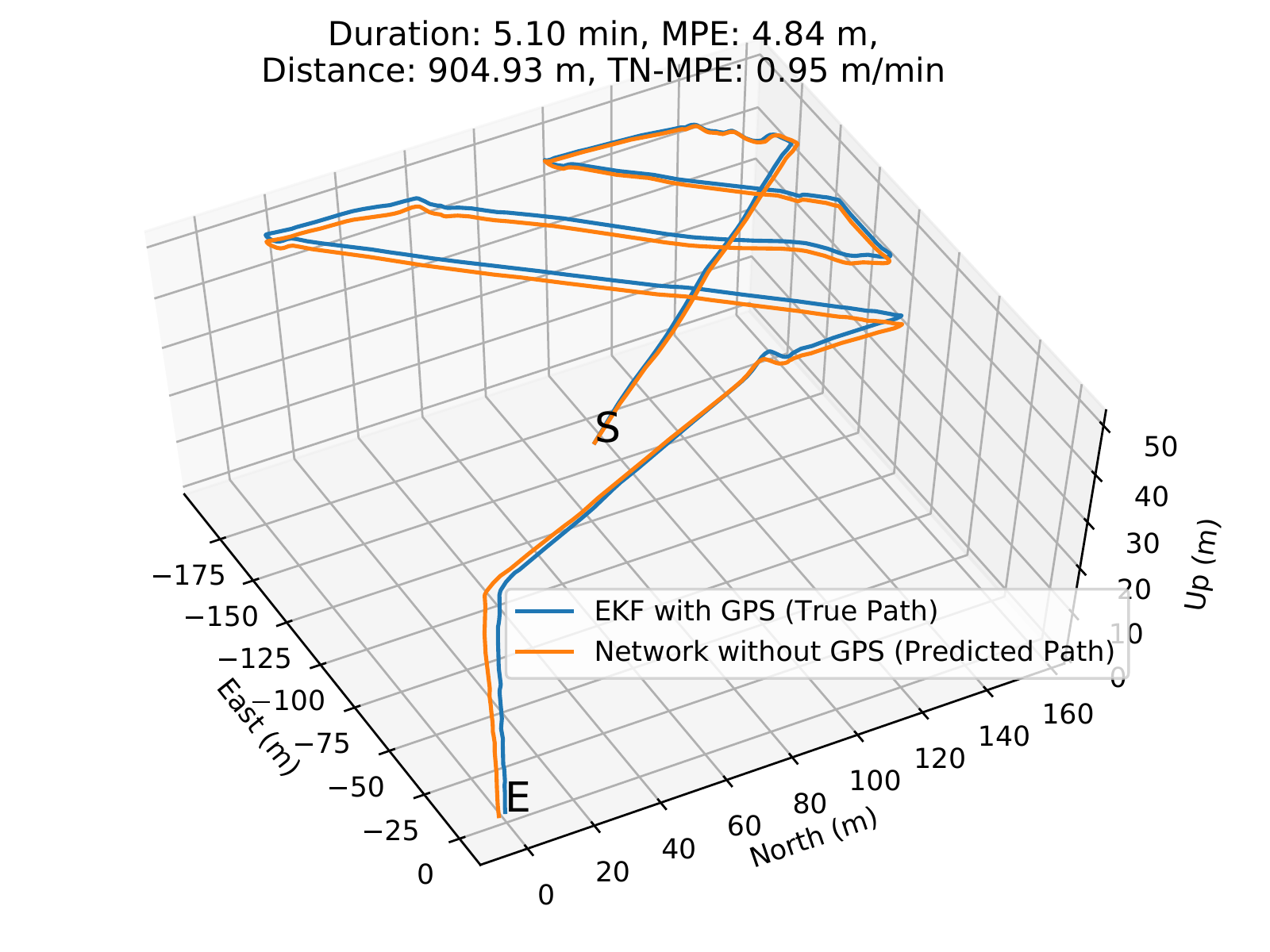}
	\caption{Validation flight path - 4}
	\label{fig:val_flight_skew_survey}
\end{figure}

Figure \ref{fig:val_flight_semi_survey}, Figure \ref{fig:val_flight_survey},
and Figure \ref{fig:val_flight_skew_survey} show three more validation
flights. Despite having the shortest flight duration, the flight in
Figure \ref{fig:val_flight_semi_survey} has the highest MPE among the three,
and has more than twice as much TN-MPE as any of them. This is because
predictions in Figure \ref{fig:val_flight_semi_survey} contain multiple
errors near the beginning. 

\subsection{Aggressive Manual Maneuvers}

A common practice in the database flights is that pilots usually start
the flight by some aggressive manual maneuvers before switching the
autopilot to ``Auto'' mode that executes the mission autonomously.
These aggressive maneuvers might excite some error sources in the
IMU that are not active in steady flights. Furthermore, the manual
sections of the flights are considerably smaller than the autonomous
ones. This, combined with the fact that these maneuvers have much
more combinations than the steady ones, means that the data capturing
their underlying mappings are \textit{rare }in the dataset. Consequently,
the network performance in manual flights is worse than that in
autonomous ones. Fortunately, manual flights do not require accurate
positioning, and these aggressive maneuvers are usually performed
within the pilot's line of sight. It is the remote autonomous missions
and steadier flights that require the proposed system the most.

\begin{figure}[h]
	\centering
	\includegraphics[width=1\linewidth]{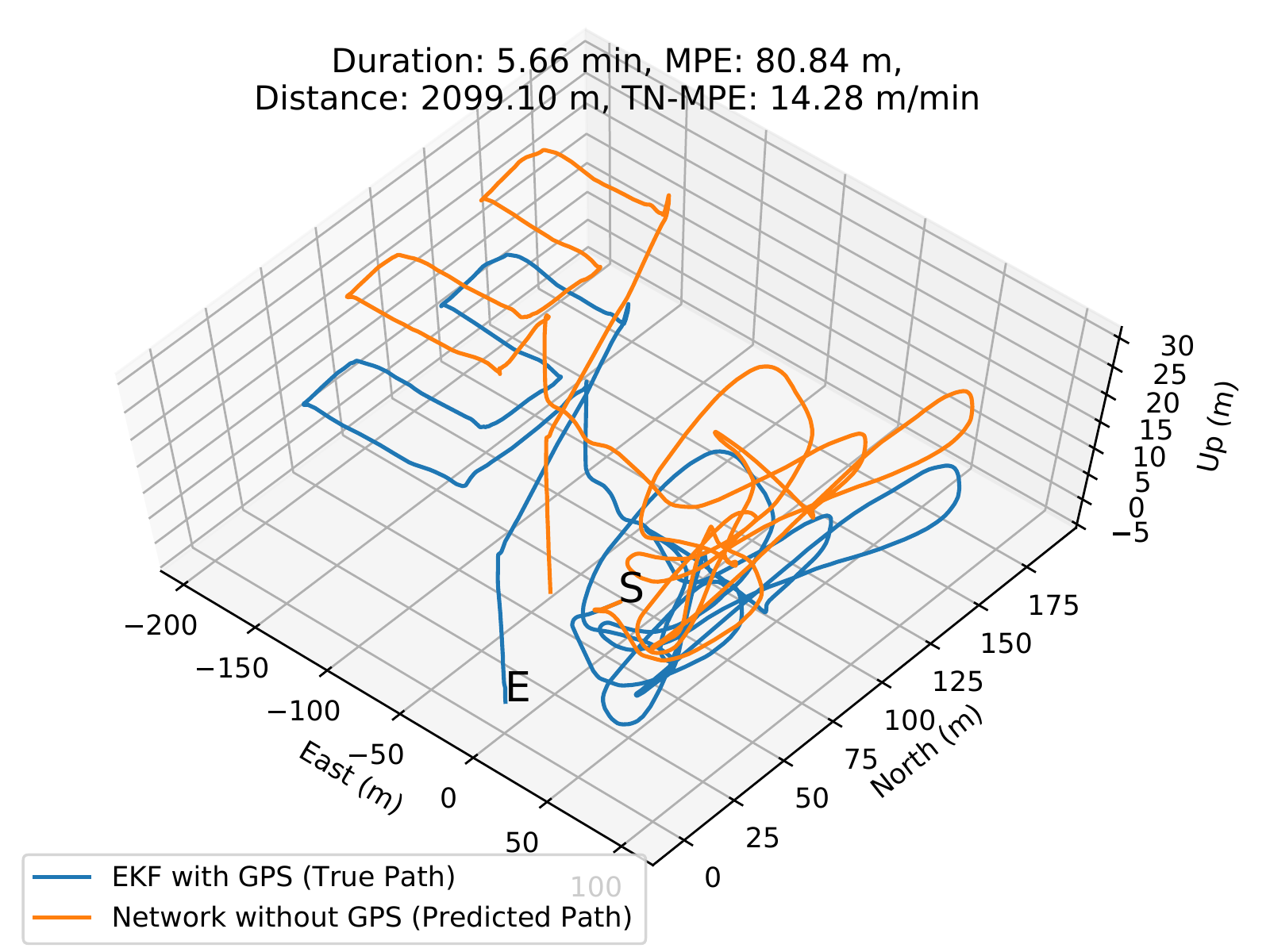}
	\caption{Validation flight with both manual	and autonomous sections}
	\label{fig:val_flight_manual_then_auto}
\end{figure}

\begin{figure}[h]
	\centering
	\includegraphics[width=1\linewidth]{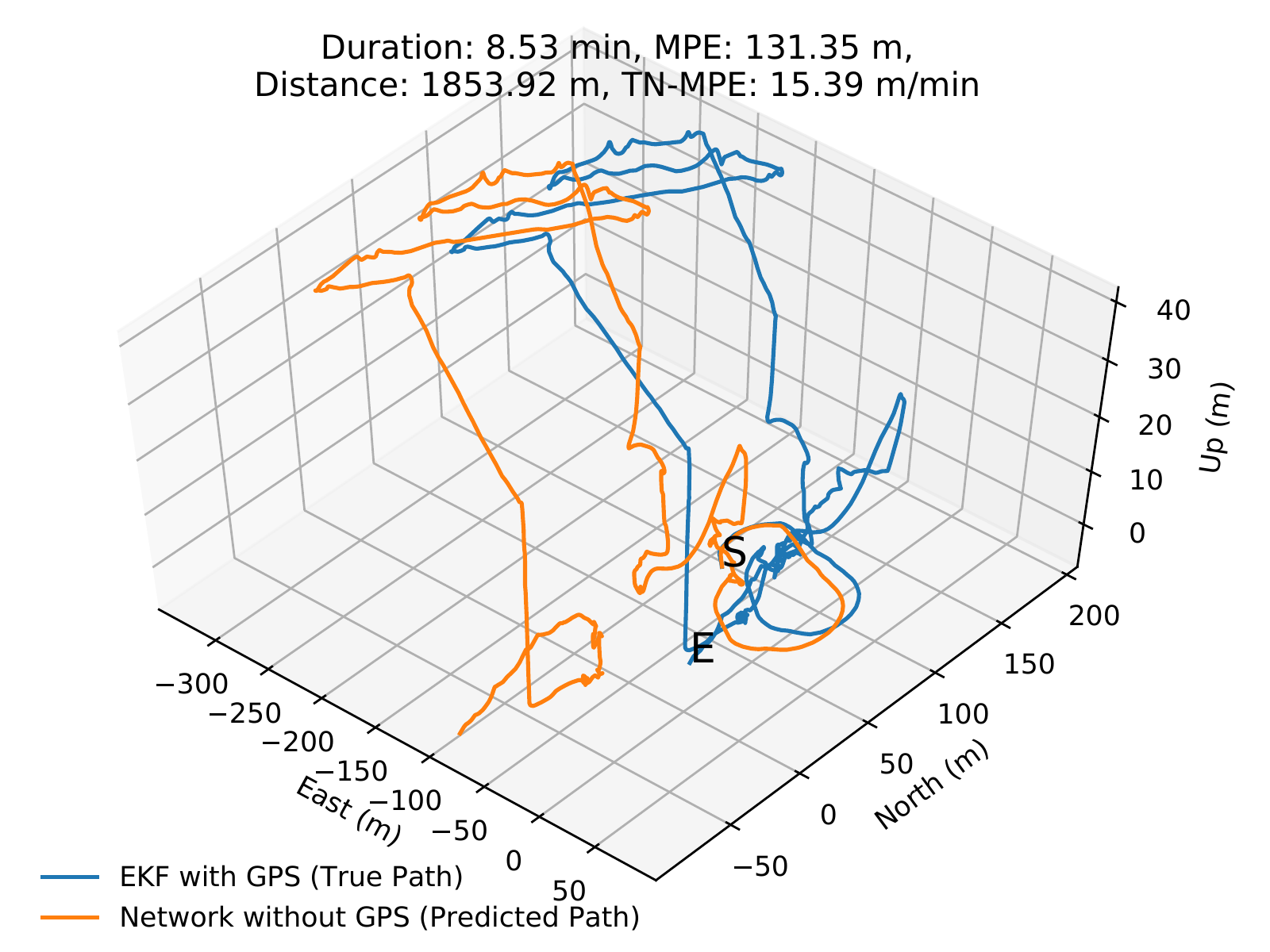}
	\caption{Validation flight with both manual	and autonomous sections -2}
	\label{fig:val_flight_manual_then_auto_2}
\end{figure}

Figure \ref{fig:val_flight_manual_then_auto} and Figure \ref{fig:val_flight_manual_then_auto_2}
show two validation flights with relatively long manual sections.
It can be noted that predictions made in these manual sections are
worse than those made in the autonomous ones. Despite these erroneous
predictions preceding the correct ones, they did not affect them.
This means that a single external reset (using an external sensor)
can remove much of the error in the latter segments. 

\begin{figure}[h]
	\centering
	\includegraphics[width=1\linewidth]{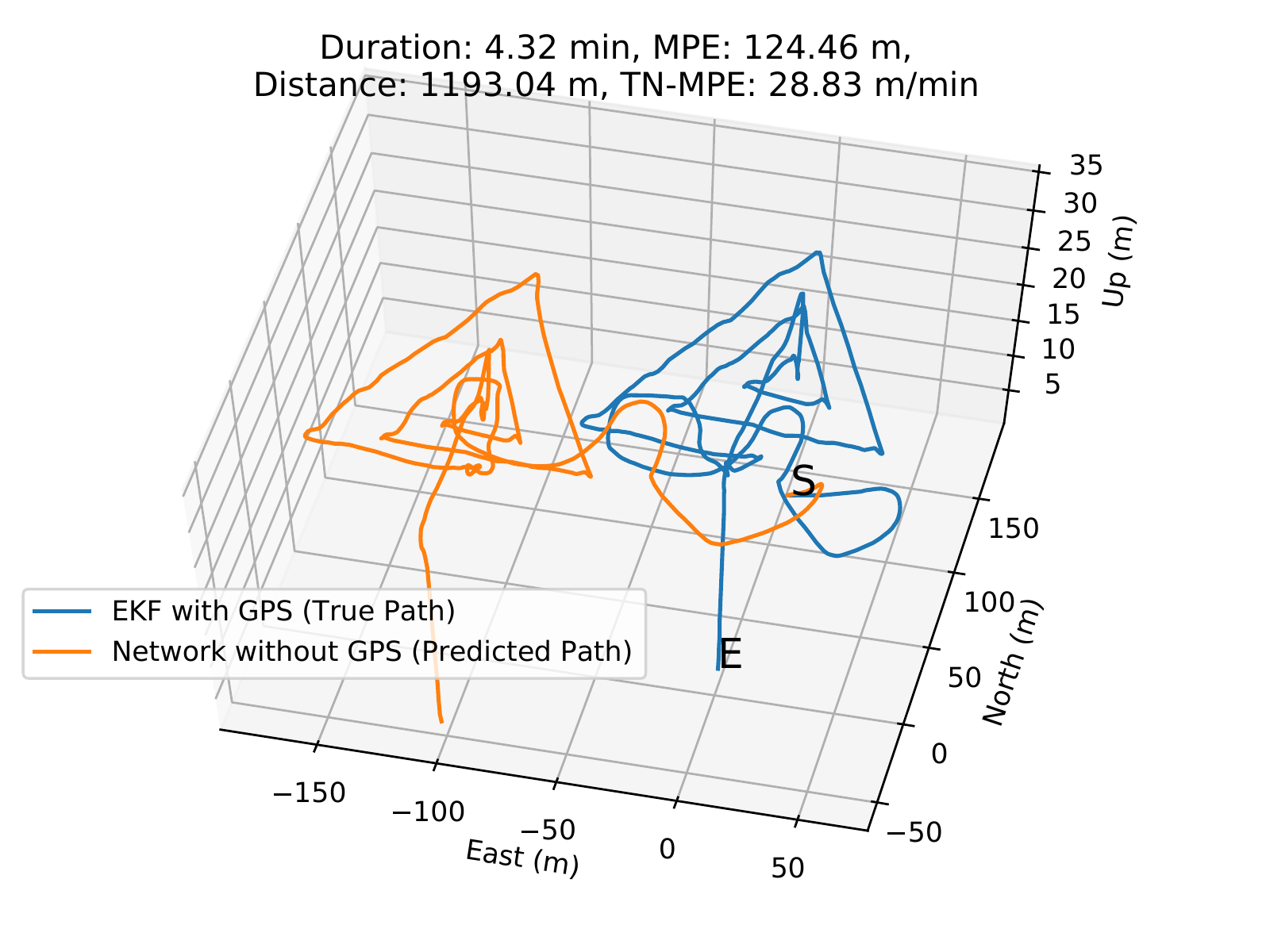}
	\caption{Validation flight with wrong predictions at the beginning}
	\label{fig:val_flight_initial_shift}
\end{figure}

Figure \ref{fig:val_flight_initial_shift} shows another flight where much
of the manual maneuvers are hidden in the first window. The ones outside
the window were enough to introduce an error that exceeds 120 meters. 

The TN-MPE index does not represent the error growth rate, despite
having rate units (m/min). This is because much of the prediction
errors shown do not grow with time. They are little-varying offsets
introduced by rare flight phenomena scattered in different and spaced
timesteps. For example, the flight in Figure \ref{fig:val_flight_initial_shift}
has a TN-MPE of 28.83 m/min, but this does not mean that if it continued
to fly one more minute, the error would have grown by 28.83 meters.
Instead, this error was introduced early in the flight and did not
change much afterward. This is also visible in Figure \ref{fig:val_flight_manual_then_auto}
and Figure \ref{fig:val_flight_manual_then_auto_2}, where much of the errors are
introduced in the beginning but do not grow later.

\subsection{Underrepresented Vehicle Types} \label{subsec:underrepresented_vehicle_types}

It was seen from Table \ref{tab:different_host_vehicle} that approximately
94\% of the flights in the dataset used a Quadrotor as a host vehicle.
The dataset is split randomly into training and validation. Only four
flights from the underrepresented vehicles fall into the validation
set, of which three used an FW vehicle and one used a VTOL. Performance
in these four flights is summarized in Table \ref{tab:underrepresented_vehicles_performance}.
The TN-MPE Rank column in Table \ref{tab:underrepresented_vehicles_performance} is the
position of a flight's TN-MPE among the entire validation set in descending
order. It can be seen that these four flights are ranked among the
five highest TN-MPEs (lowest accuracy) in the validation set.

\begin{table}[h]
	\caption{Validation performance in underrepresented	vehicle types}
	\centering
	\setlength{\tabcolsep}{1pt}	
	\begin{tabular}{| M{0.15\linewidth} | M{0.15\linewidth} |  M{0.15\linewidth}| M{0.15\linewidth} | M{0.15\linewidth} |  M{0.15\linewidth}|}
		\hline 
		Flight & Duration (min) & Distance (m) & MPE (m) & TN-MPE (m/min) & TN-MPE Rank\\
		\hline 
		\hline 
		FW \#1 & 0.73 & 488.73 & 103.72 & 142.08 & $1^{st}$\\
		\hline 
		FW \#2 & 6.52 & 5424.57 & 871.54 & 133.74 & $2^{nd}$\\
		\hline 
		FW \#3 & 2.6 & 2062.32 & 289.82 & 111.47 & $4^{th}$\\
		\hline 
		VTOL & 5.53 & 3620.59 & 551.58 & 99.8 & $5^{th}$\\
		\hline 
	\end{tabular}
	\label{tab:underrepresented_vehicles_performance}
\end{table}

\begin{figure}[h]
	\centering
	\includegraphics[width=1\linewidth]{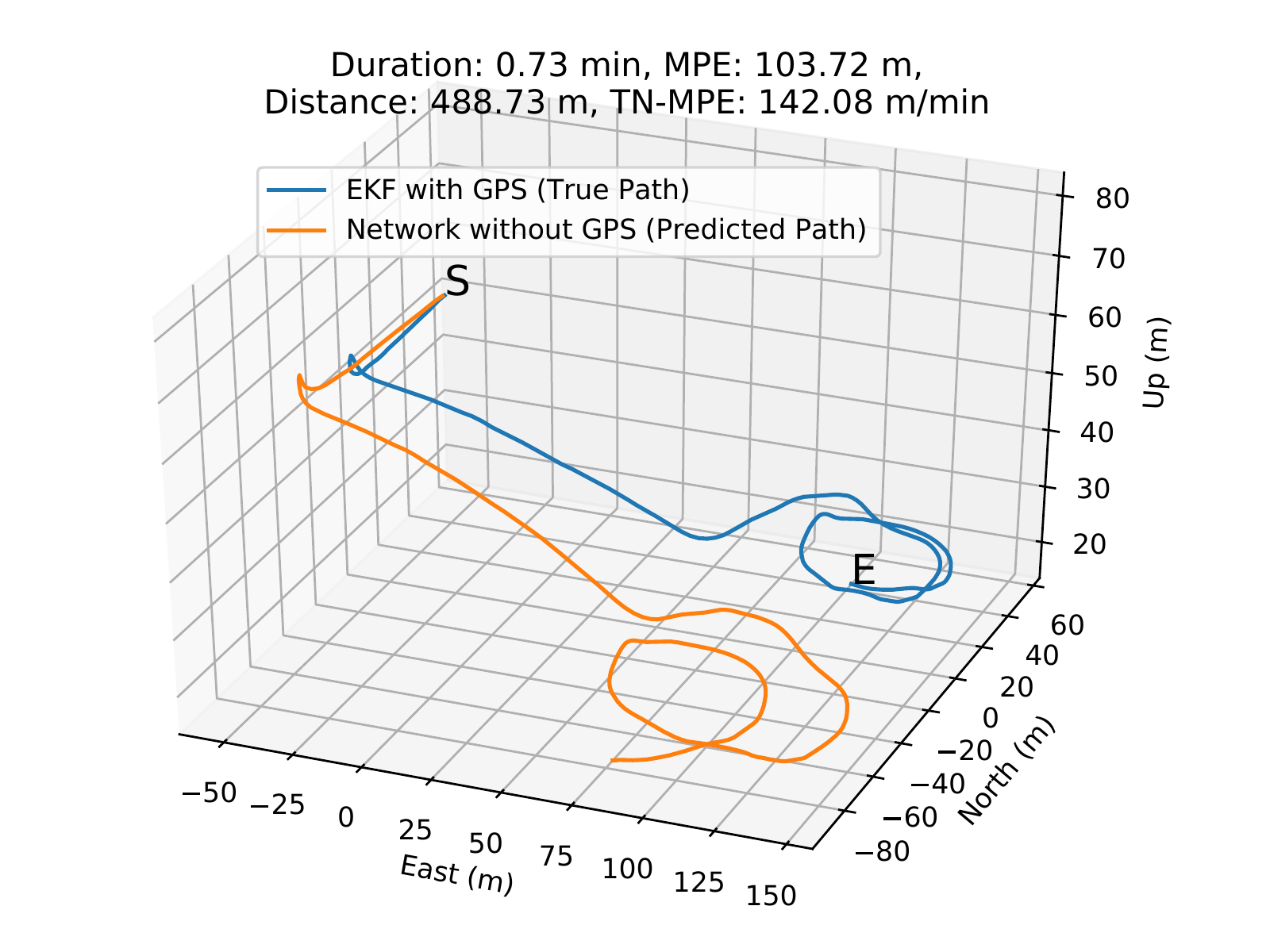}
	\caption{FW Validation Flight \# 1}
	\label{fig:val_flight_FW}
\end{figure}

Figure \ref{fig:val_flight_FW} shows the FW validation flight \# 1.
Despite the shorter than a minute duration, positioning error grows
beyond a hundred meters. The flight does not have aggressive manual
maneuvers either.

\begin{figure}[h]
	\centering
	\includegraphics[width=1\linewidth]{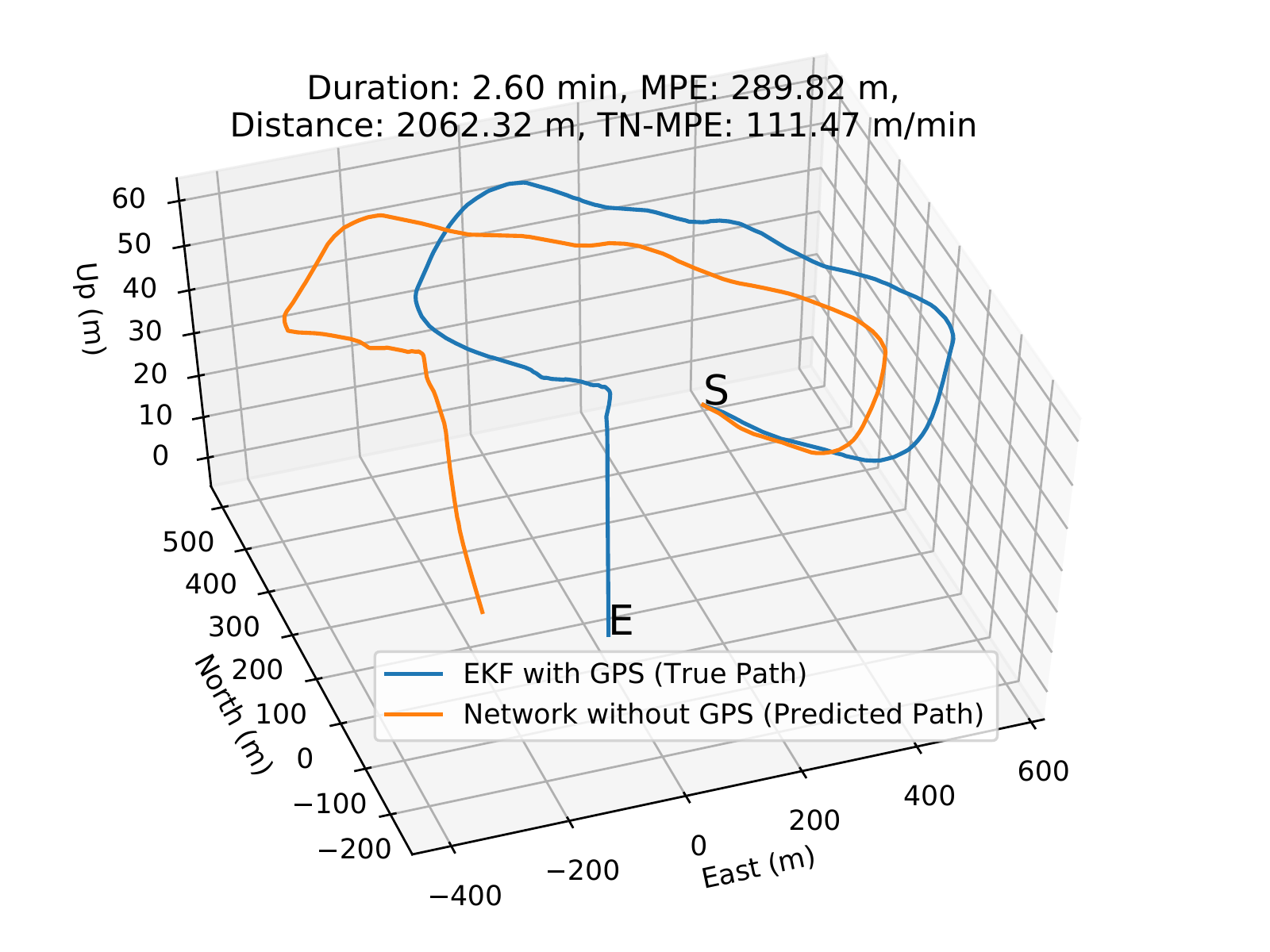}
	\caption{VTOL validation flight}
	\label{fig:val_flight_VTOL}
\end{figure}

The same observation applies to the VTOL validation flight in Figure \ref{fig:val_flight_VTOL}.
This, too, is a short flight with smooth maneuvers; still, positioning
error grows to almost 300 meters in less than three minutes.

Some factors contributing to the relatively high TN-MPE with underrepresented
vehicles are:
\begin{enumerate}
	\item The network learns to compensate for the Magnetometer errors at some
	point in the prediction process. These errors are affected mainly
	by the magnetic field of the vehicle's motors, which varies from one
	vehicle type to another. For example, an FW drone usually has only
	one motor, but the Quadrotor has four, and the Standard VTOL has five. 
	\item The network may be learning to compensate for vibrations, which
	vary from one vehicle type to another. 
	\item These flights have relatively high average velocities, the average
	velocities of the flights in Table \ref{tab:underrepresented_vehicles_performance} range
	from 10.91 m/s to 13.87 m/s. The mean value of average velocities
	in the validation set is only 4.66 m/s.
\end{enumerate}

Other than these worst five flights, the sixth-highest TN-MPE in the
validation flights is only 52.38 m/min, followed by 38.94 m/min. This
explains the large gap between the mean and median values for the
TN-MPE in Table \ref{tab:validation_performance}.

\subsection{Velocity Predictions}

The network described in Table \ref{tab:network_architecture_and_hyperparam} predicts
both position and velocity differences (increments) independently,
so errors in some direction in velocity are not necessarily reflected
in position errors in the same direction. Another method to calculate
the velocity is to divide the position differences calculated by the
network by the time step. Since the time step is fixed to 0.2 seconds,
any predicted position difference can be divided by 0.2 to get the
velocity in the same direction. Not only does this reduce the network
size and computations, but it also improves the accuracy of the calculated
velocity. 

\begin{figure}[h]
	\centering
	\includegraphics[width=1\linewidth]{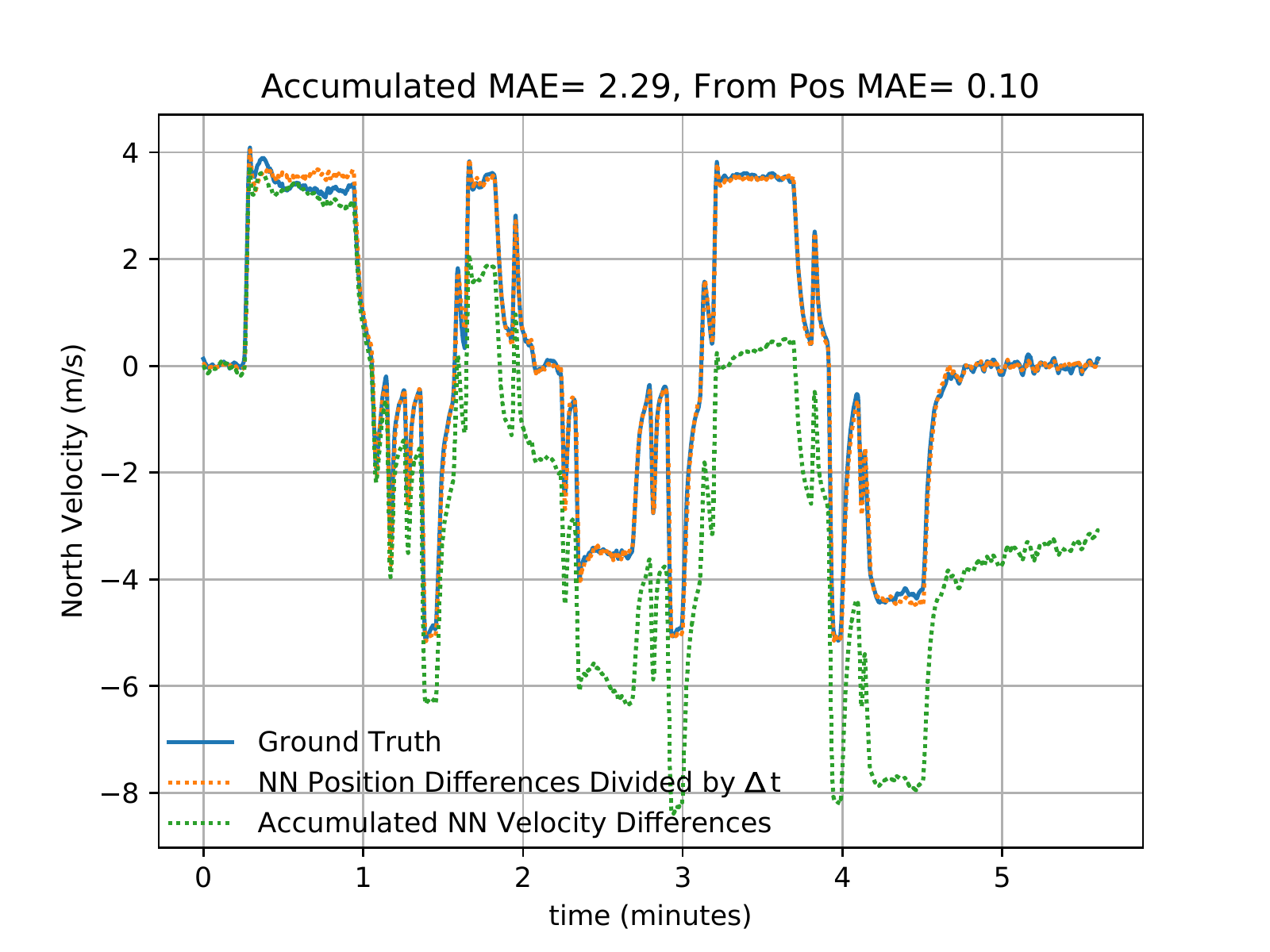}
	\caption{North velocity component of a validation flight as predicted by the network vs. as calculated from
		position differences}
	\label{fig:north_velocity}
\end{figure}

\begin{figure}[h]
	\centering
	\includegraphics[width=1\linewidth]{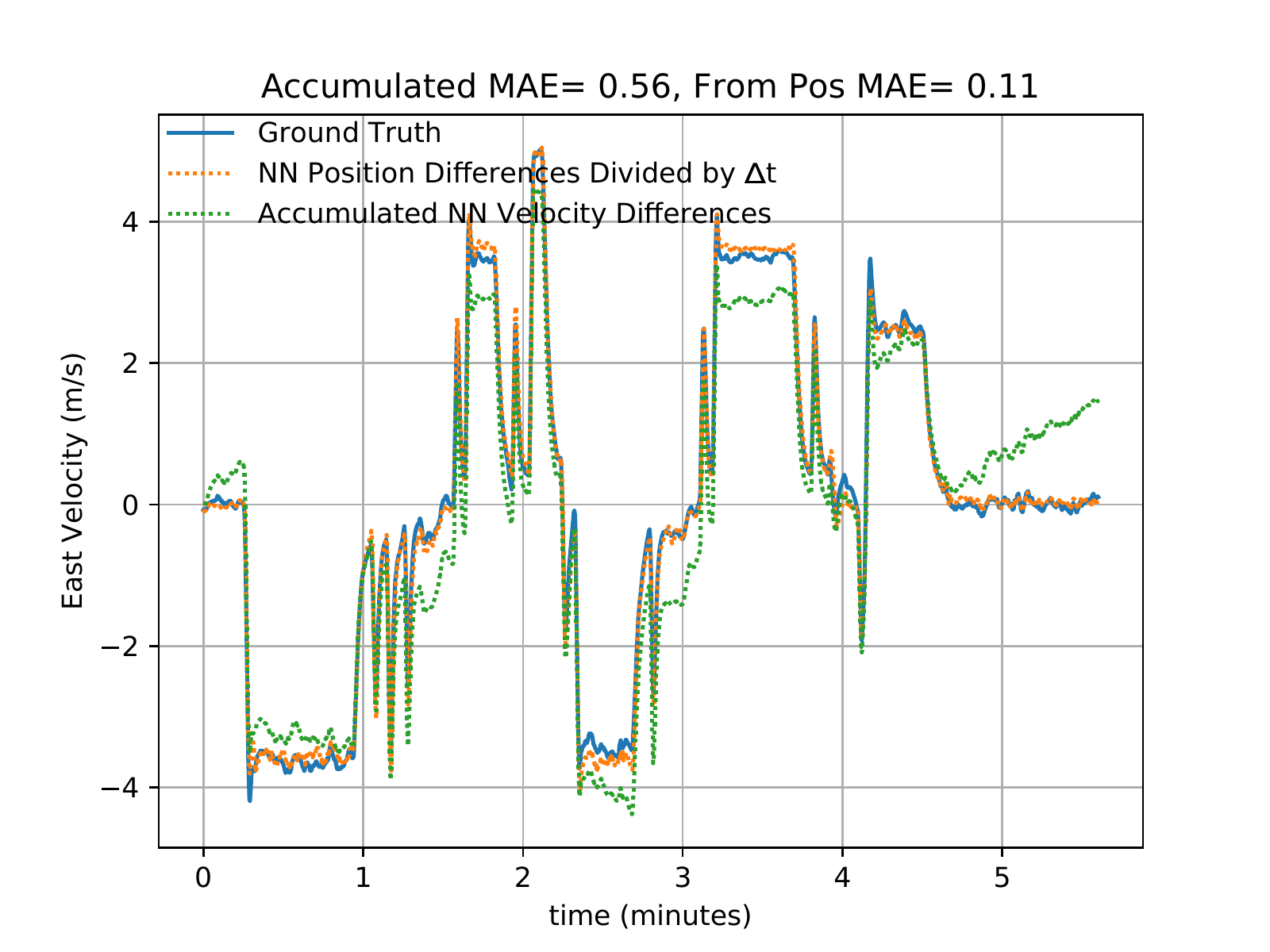}
	\caption{East velocity component of a validation flight as predicted by the network vs. as calculated from position
		differences}
	\label{fig:east_velocity}
\end{figure}

\begin{figure}[h]
	\centering
	\includegraphics[width=1\linewidth]{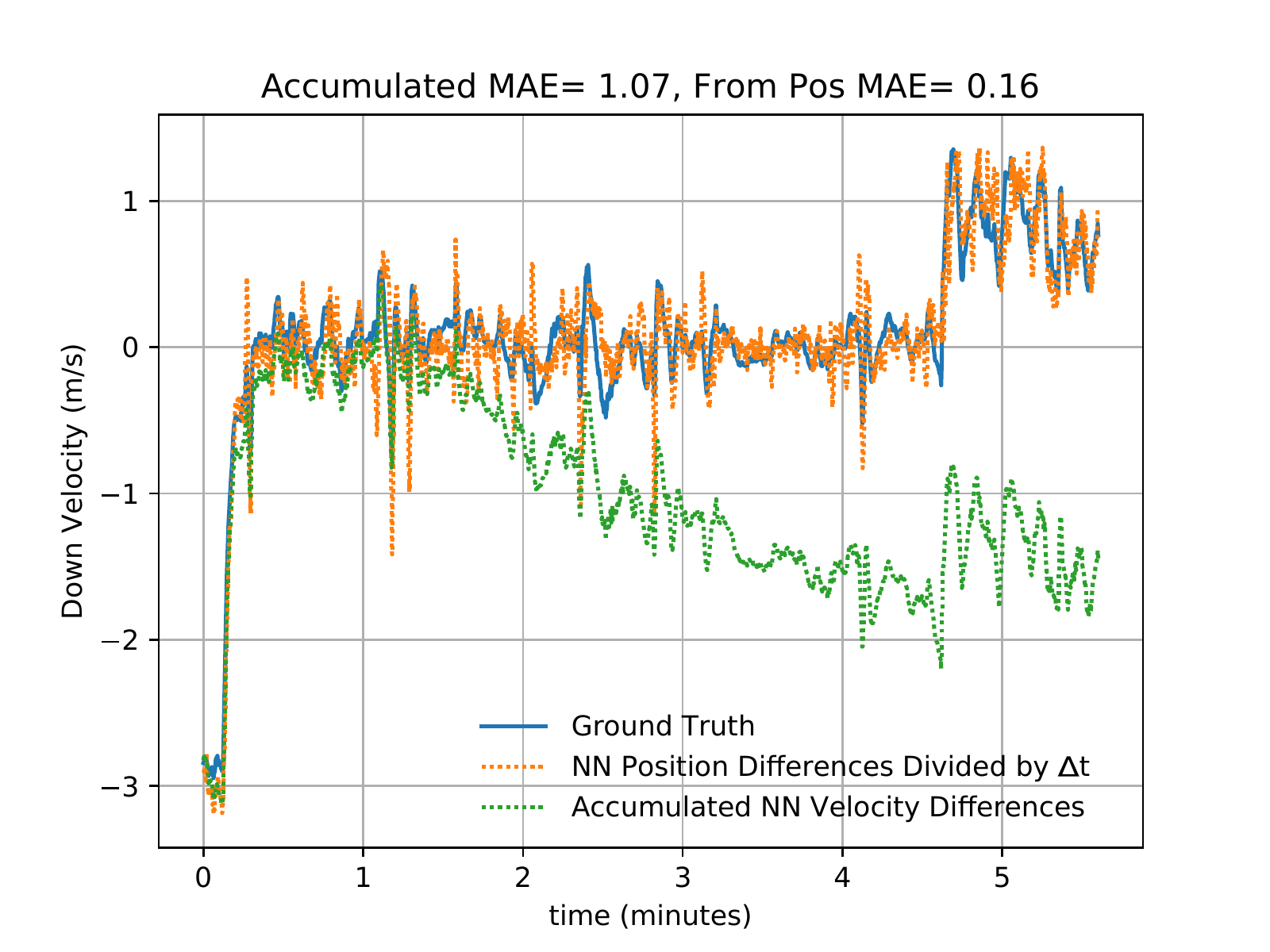}
	\caption{Down velocity component of a validation flight as predicted by the network vs. as calculated from position
		differences}
	\label{fig:down_velocity}
\end{figure}

Figure \ref{fig:north_velocity}, Figure \ref{fig:east_velocity},
and Figure \ref{fig:down_velocity} compare the two methods in the
three velocity components in a validation flight.

\subsection{Attitude Predictions}

\begin{figure}[h]
	\centering
	\includegraphics[width=1\linewidth]{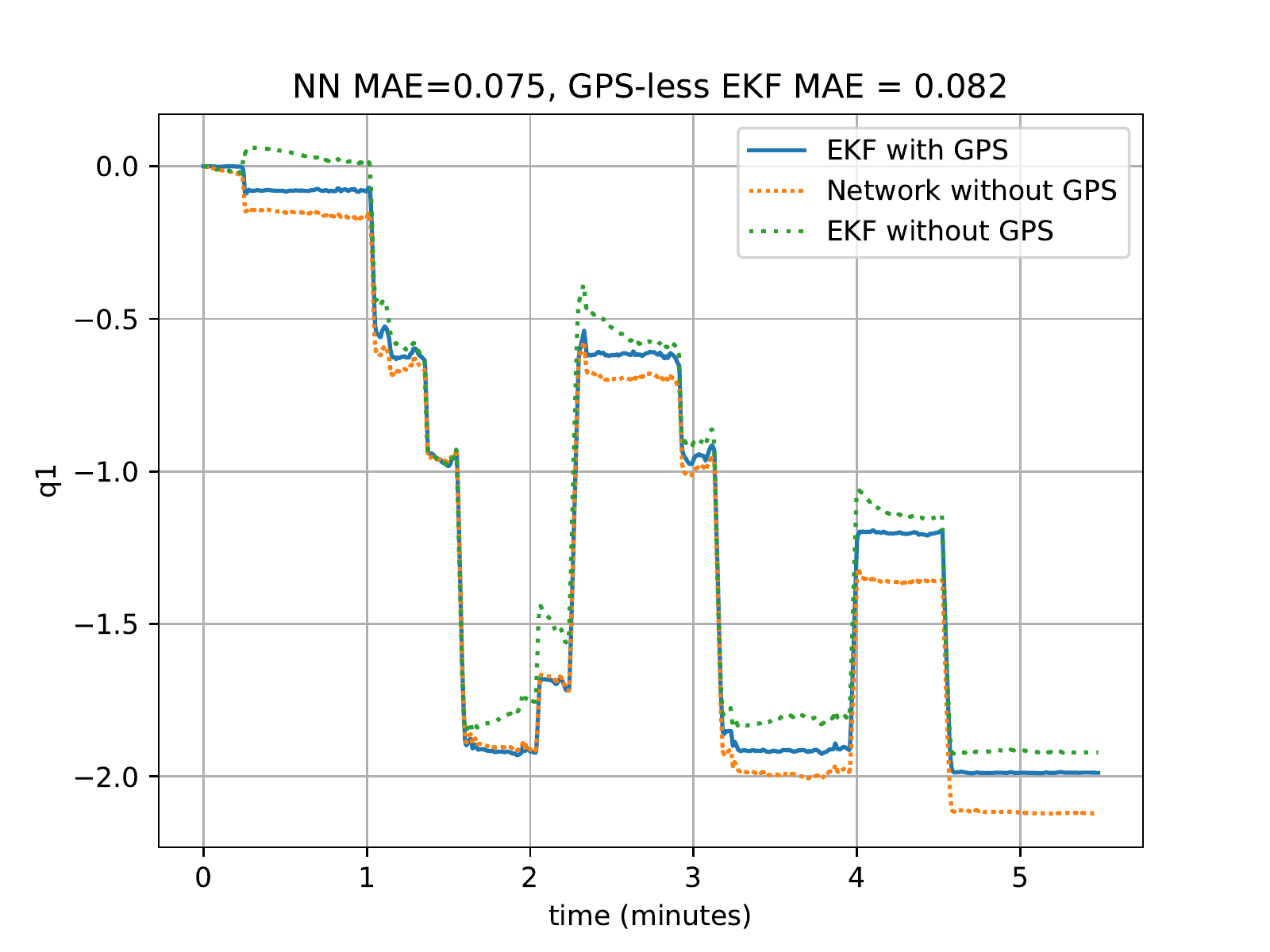}
	\caption{Attitude quaternions - q1}
	\label{fig:quaternions_q1}
\end{figure}

\begin{figure}[h]
	\centering
	\includegraphics[width=1\linewidth]{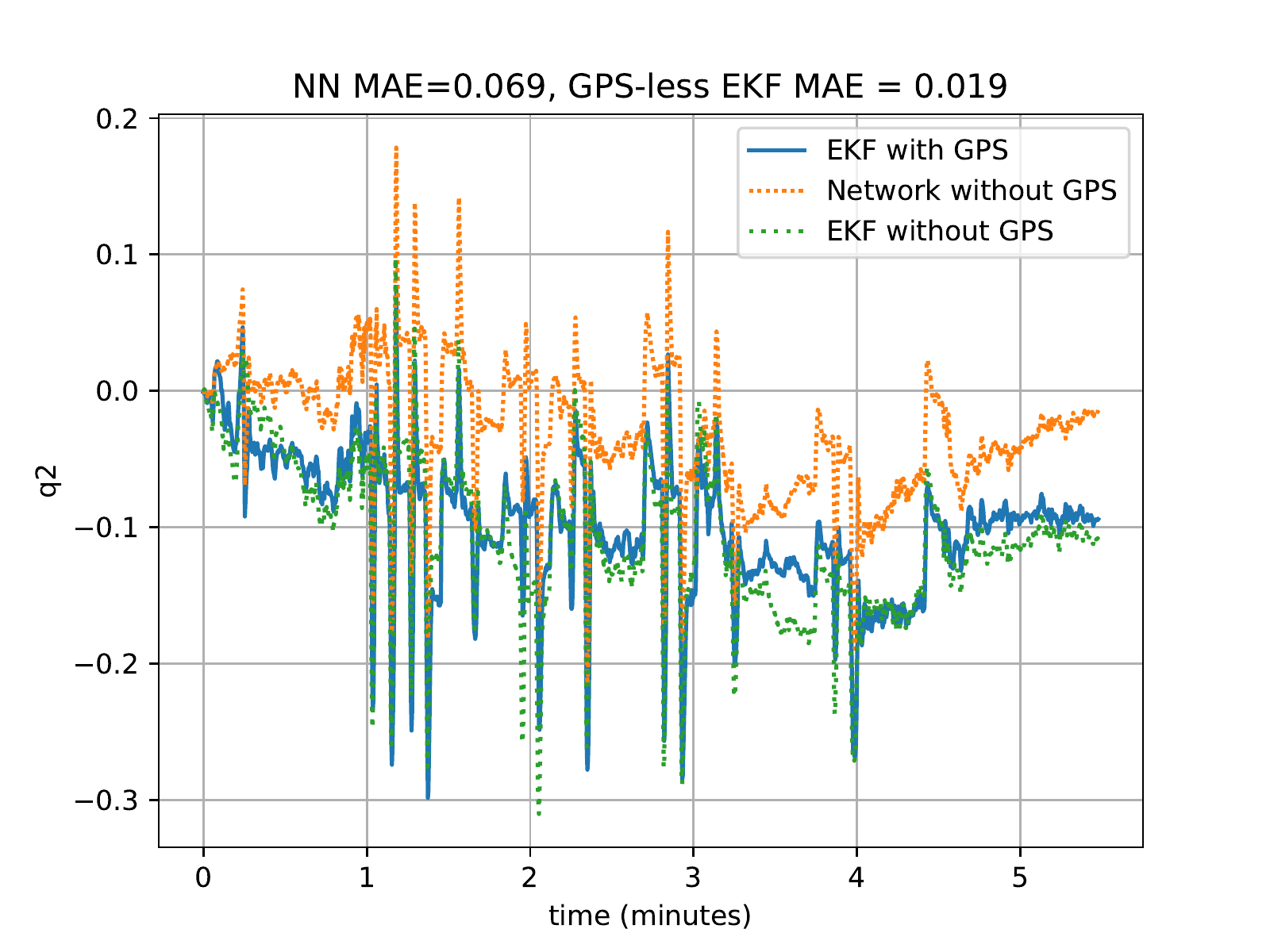}
	\caption{Attitude quaternions - q2}
	\label{fig:quaternions_q2}
\end{figure}

\begin{figure}[h]
	\centering
	\includegraphics[width=1\linewidth]{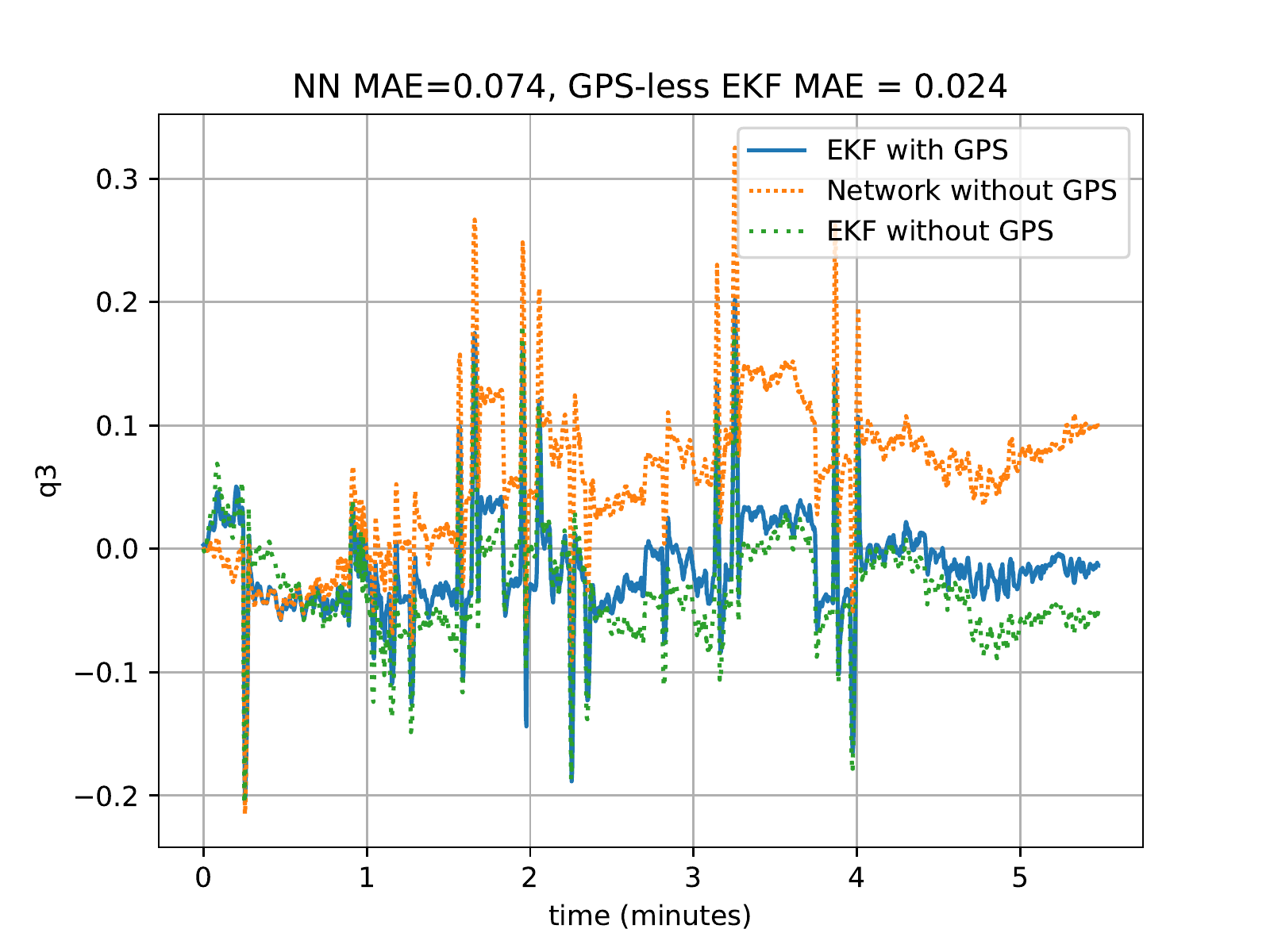}
	\caption{Attitude quaternions - q3}
	\label{fig:quaternions_q3}
\end{figure}

\begin{figure}[h]
	\centering
	\includegraphics[width=1\linewidth]{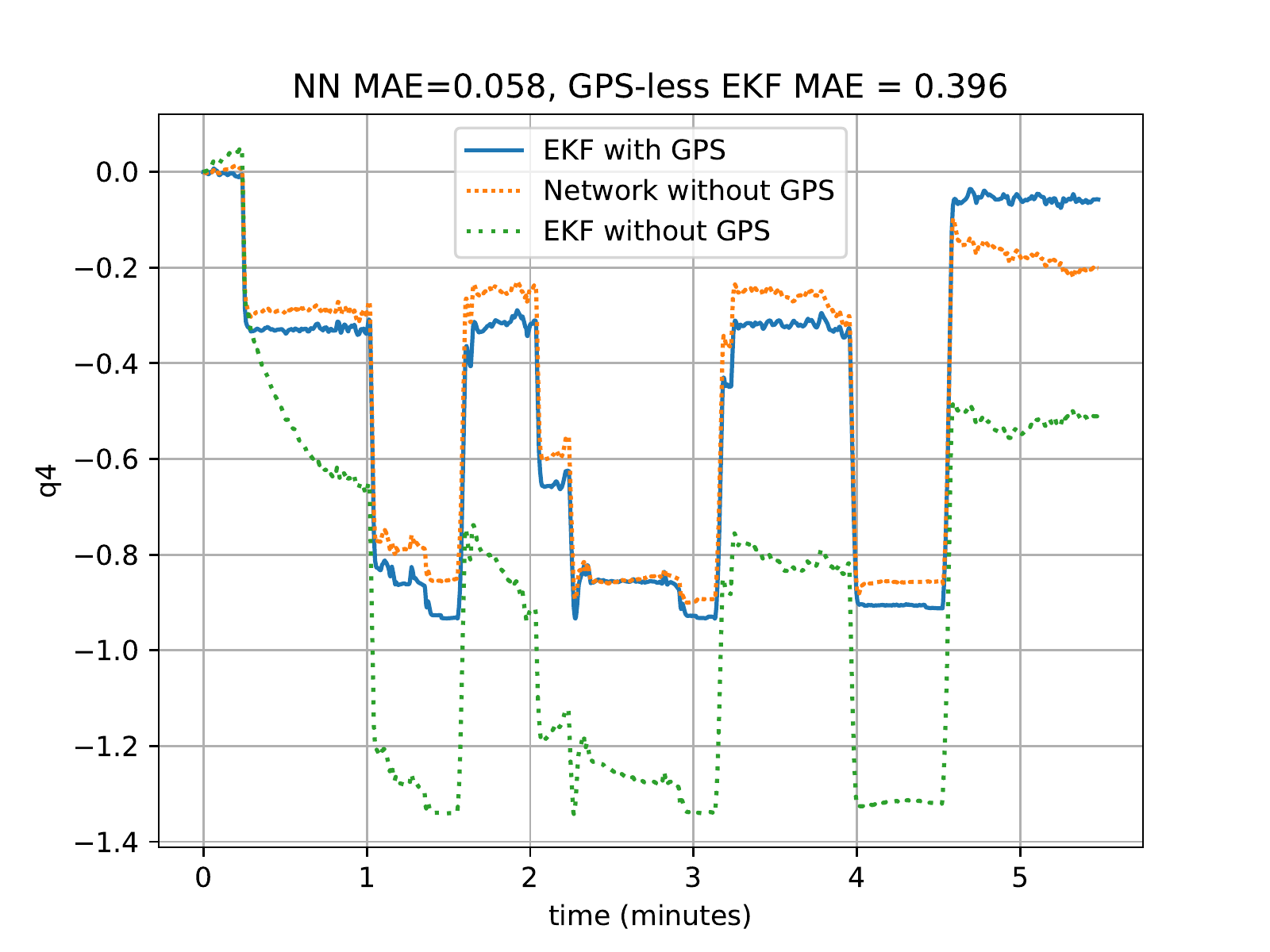}
	\caption{Attitude 1uaternions - q4}
	\label{fig:quaternions_q4}
\end{figure}

The PX4 EKF2 estimates the attitude expressed in quaternions. These
attitude estimates do not diverge when the GPS signal is lost, because
Gyroscope and Accelerometer can predict reliable roll and pitch angles,
and the Magnetometer helps correct the yaw angle. Another advantage
of the EKF's attitude is that it is produced at a relatively high
rate of 84 Hz. The network, on the other hand, is trained using 5
Hz logs, so it is limited by this rate. Some trials where the network
predicted the attitude along with velocity and position were performed.
Figure \ref{fig:quaternions_q1}, Figure \ref{fig:quaternions_q2},
Figure \ref{fig:quaternions_q3} and Figure \ref{fig:quaternions_q4}
show the attitude as predicted by the network and as estimated by
the EKF when GPS is lost. The EKF errors here result from the
numerical complications caused by the massively diverging position
and velocity estimates. In the PX4 code, this is cured by feeding
zero position and velocity measurements to the EKF when the GPS fix
is lost to preserve the numerical stability of other estimated quantities
like the attitude.

These reasonable estimates produced by the GPS-less EKF are the reason why
the final network design does not predict attitude. Adding more elements
to the labels vector makes the network more complex and requires more
time for both training and inference. Requiring more labels from the
network also diverts some focus from the position and velocity calculations
and reduces their performance.

\section{Field Testing and Real-time Inference}

Inference time is the time taken by the network to predict one labels
vector. That is, given a window of sensor measurements, how much time
does the network need to predict the corresponding change of position
and velocity at the end of this window? Since the network is trained
to predict position at 5 Hz, inference time must be less than 200
milliseconds. Inference time is determined by three factors, the embedded
hardware on which the network is deployed, the length of the input window, and the size of the network
itself. Larger networks achieve higher accuracy but require stronger
hardware. Increasing the hardware capabilities comes with increasing size, power consumption, weight, and cost. This is why a compromise
is usually made between the network size and its accuracy. 

\begin{figure}[h]
	\centering
	\includegraphics[width=1\linewidth]{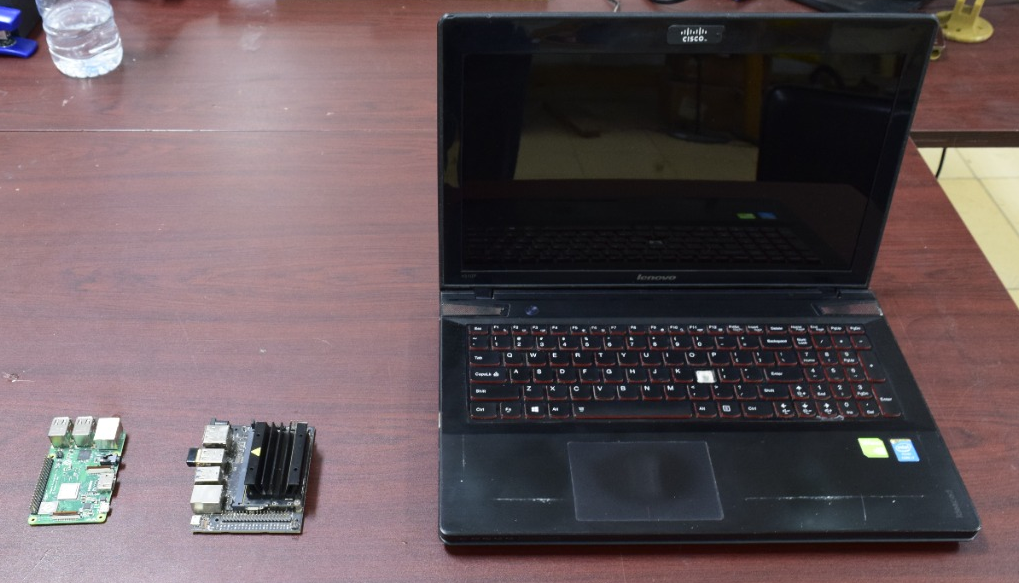}
	\caption{Companion computers tested	to run the network in real-time, from left to right (Raspberry Pi
		3 B+, Jetson Nano, Lenovo Laptop)}
	\label{fig:companion_computers}
	
\end{figure}

The Pixhawk itself cannot run the heavy NN calculations. This is why
a companion computer is usually used alongside it to perform the intensive
computations. A smaller version of the network, with only 50 steps
window and three labels (position only) was deployed on different companion
computers to measure inference time. On a Raspberry Pi 3 B+ \cite{FOUNDATION2021},
inference time is 840 ms, on Nvidia's Jetson Nano \cite{NVIDIA2020},
inference time is 200 ms, and on a core i7 laptop \cite{Lenovo2021},
inference time is 50 ms. Faster inferences can be achieved with some
sacrifice in accuracy. Figure \ref{fig:companion_computers} shows
the tested companion computers.

\begin{figure}[h]
	\centering
	\includegraphics[width=1\linewidth]{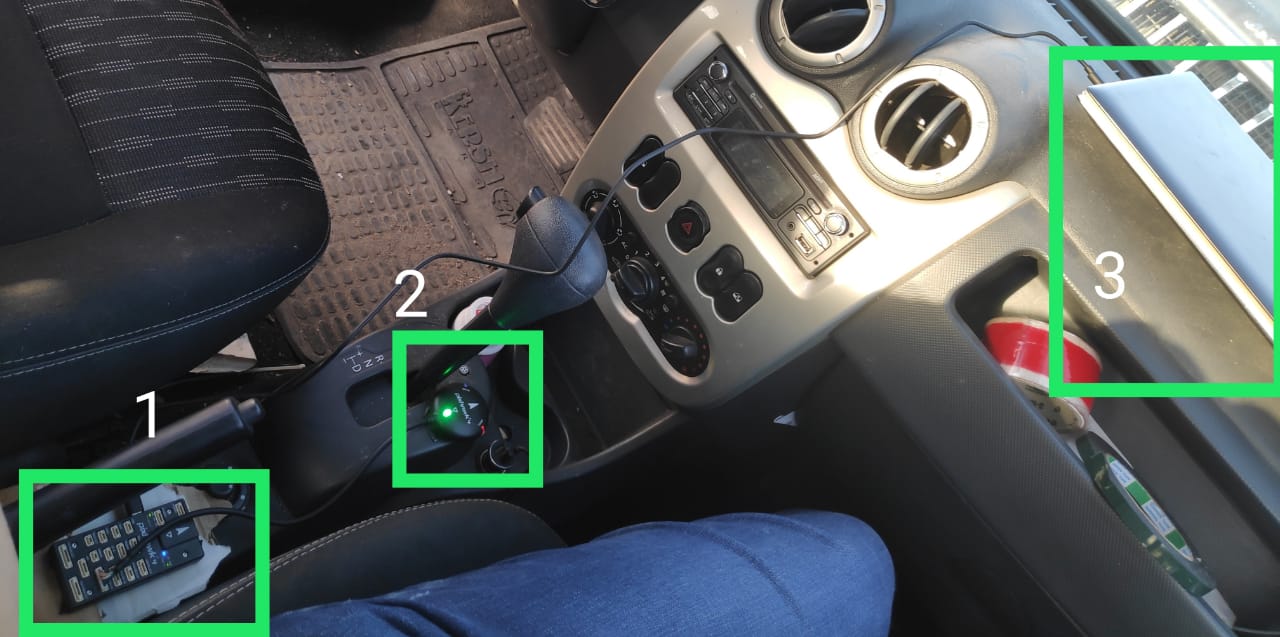}
	\caption{Hardware Setup in a Car}
	\label{fig:car_hardware_setup}
\end{figure}

To test the real-time performance, a Pixhawk4 was mounted to a car
and the laptop running the ground station was also used as the companion
computer. The configuration only requires the Pixhawk itself, a GPS
receiver (to compare to ground truth), and a laptop, labeled 1,2 and
3 respectively in Figure \ref{fig:car_hardware_setup}. The laptop provides
power to the Pixhawk, so the battery and power module are not needed. The
same cable used to power the Pixhawk from the laptop also works as
a data link; this eliminates the need for a telemetry module.

As explained in section \ref{subsec:underrepresented_vehicle_types}, changing
the host vehicle dramatically reduces the network performance. Ground vehicles
are not just underrepresented in the training dataset, they are not
represented at all. This meant that new logs had to be collected using
cars. A total of 34 ground trips were performed using two cars, with
a combined duration of about 6.5 hours. Only five trips were held for
validation (about half an hour) and the rest were used for training. 

A popular technique used in Deep Learning to compensate for the small
dataset sizes is Transfer Learning \cite{Tan2018,Weiss2016}. Transfer
learning means that a network trained for some task can be used as
a starting point to train another network for another task. For example,
the trained weights can be used as the initial weights for a new task,
instead of randomly initializing the weights. This should be done
carefully to avoid the loss of the relations learned from the large
dataset and at the same time accommodate the new relations introduced
by the smaller set.

The same concept can be applied to different vehicle types, for example,
a network can be pretrained on Quadrotors, then the knowledge is transferred
to a network that works with fixed-wing aircraft, or in this case,
a car.

\begin{figure}[h]
	\centering
	\includegraphics[width=1\linewidth]{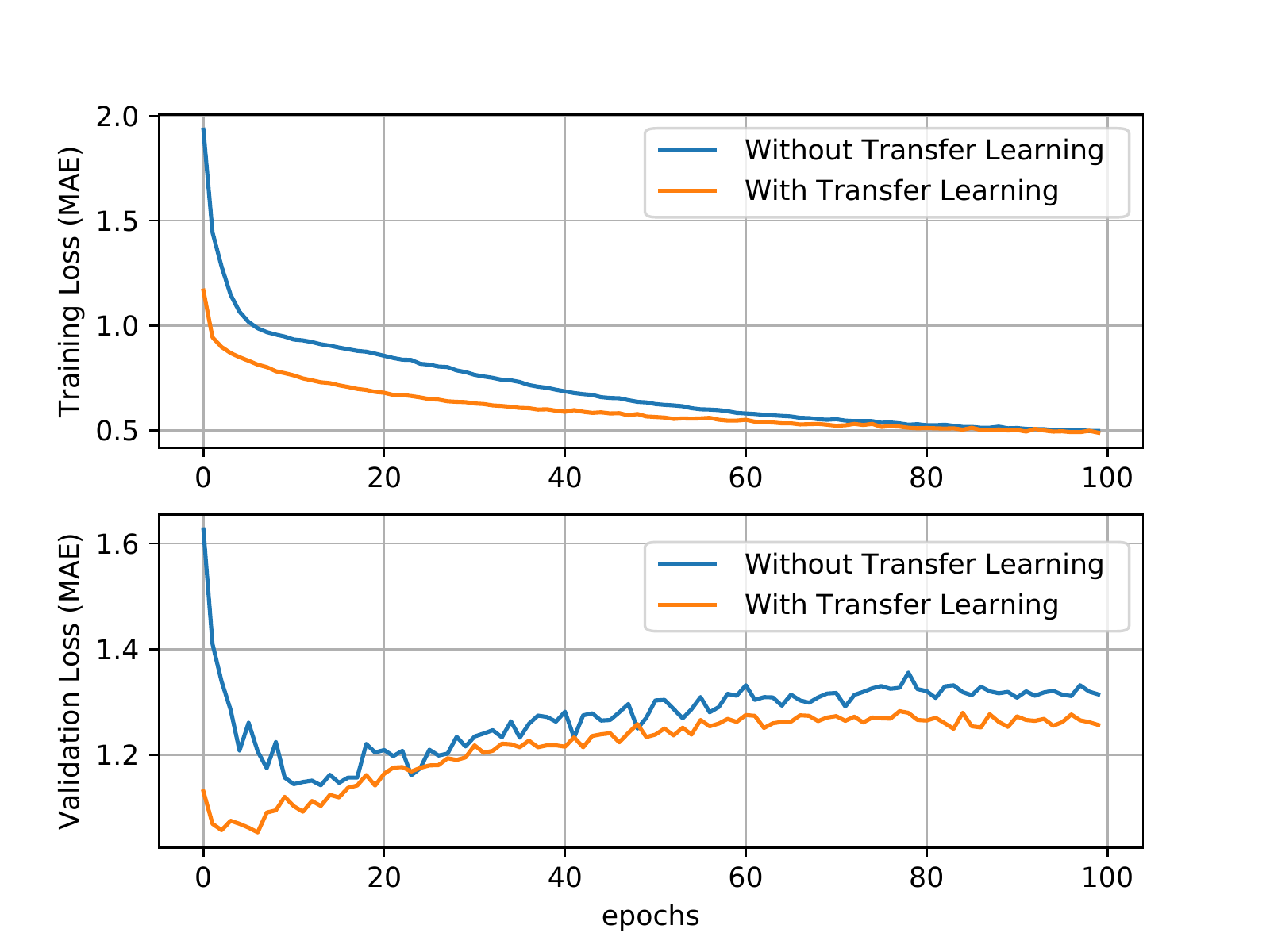}
	\caption{Effect of transfer learning}
	\label{fig:benefit_of_transfer_learning}
\end{figure}

Three experiments were conducted on the light network. The first experiment
was to train it on the drones dataset from Table \ref{tab:different_host_vehicle}
then to use it without any modifications on the cars dataset. The
second experiment trains it from scratch on the cars dataset
for 100 epochs without utilizing transfer learning. The last experiment
is to pretrain it on the drones dataset and then to retrain it on the
cars dataset, which is a form of transfer learning. Figure \ref{fig:benefit_of_transfer_learning}
shows the effect of utilizing transfer learning. It is seen that fresh
training starts overfitting after 17 epochs, while retraining overfits
earlier (around five epochs). This is the point where the network starts
to \textit{forget} the relations it has learned from the drones dataset.
It is also noticeable that the validation loss is reduced when transfer
learning is applied. 

A comparison between the three experiments is shown in Table \ref{tab:car_transfer_learning}.
The validation results listed utilize early stopping, i.e., overfitting
is avoided and the weights used are not the final overfitted weights.

\begin{table}[h]
	\caption{Results on the car validation dataset with and without transfer learning}
	\centering
	\setlength{\tabcolsep}{0.5pt}
	\begin{tabular}{|M{0.33\linewidth}|M{0.18\linewidth}|M{0.16\linewidth}|M{0.3\linewidth}|}
		\hline 
		& Trained on drones only & Trained on cars only & Trained on drones, retrained on cars\\
		\hline 
		\hline 
		Median MPE (m) & 4250.92 & 732.12 & 520.19\\
		\hline 
		Mean MPE (m) & 7261.254 & 1133.324 & 707.854\\
		\hline 
		Median TN-MPE (m/s) & 799.04 & 99.41 & 72.99\\
		\hline 
		Mean TN-MPE (m/s) & 817.474 & 181.124 & 97.222\\
		\hline 
	\end{tabular}
	\label{tab:car_transfer_learning}
\end{table}

Table \ref{tab:car_validation} shows the results on the validation
car trips obtained when transfer learning is applied.

\begin{table}[h]
	\caption{Validation performance on the car dataset}
	\setlength{\tabcolsep}{1pt}
	\begin{tabular}{|M{0.4\linewidth}|M{0.14\linewidth}|M{0.14\linewidth}|M{0.14\linewidth}|M{0.14\linewidth}|}
		\hline 
		Trip Number & 1 & 2 & 3 & 4\\
		\hline 
		\hline 
		Duration (minutes) & 5.32 & 5.58 & 9.2 & 11.35\\
		\hline 
		Traveled Distance (Kilometers) & 3.314 & 1.385 & 8.55 & 13.79\\
		\hline 
		NN MPE (Kilometers) & 0.52 & 0.136 & 0.388 & 1.282\\
		\hline 
		TN-MPE (m/min) & 72.99 & 24.42 & 56.52 & 113.05\\
		\hline 
		NN MPE as a Percent of Traveled Distance & 11.72 \% & 9.83 \% & 6.1 \% & 9.3 \%\\
		\hline 
		GPS-less EKF MPE (Kilometers) & 24.184 & 25.752 & 203.77 & 421.99\\
		\hline 
	\end{tabular}
	\label{tab:car_validation}
\end{table}

It is seen from Table \ref{tab:car_validation} that the TN-MPE
values of the car trips are lower than those of the FW and VTOL flights
mentioned in Table \ref{tab:underrepresented_vehicles_performance}. This is because
the collected car trips are longer than those of FW and VTOL, and
because of the retraining, which gave more attention to this small
set.

The accuracy on the car dataset is smaller than that of the flight
dataset because the latter is much larger. If higher accuracy is needed
on ground vehicles, then more logs must be collected. However,
the purpose of car testing in this context is mainly to address the
challenges associated with real-time inference (edge computing). 

\begin{figure}[h]
	\centering
	\includegraphics[width=1\linewidth]{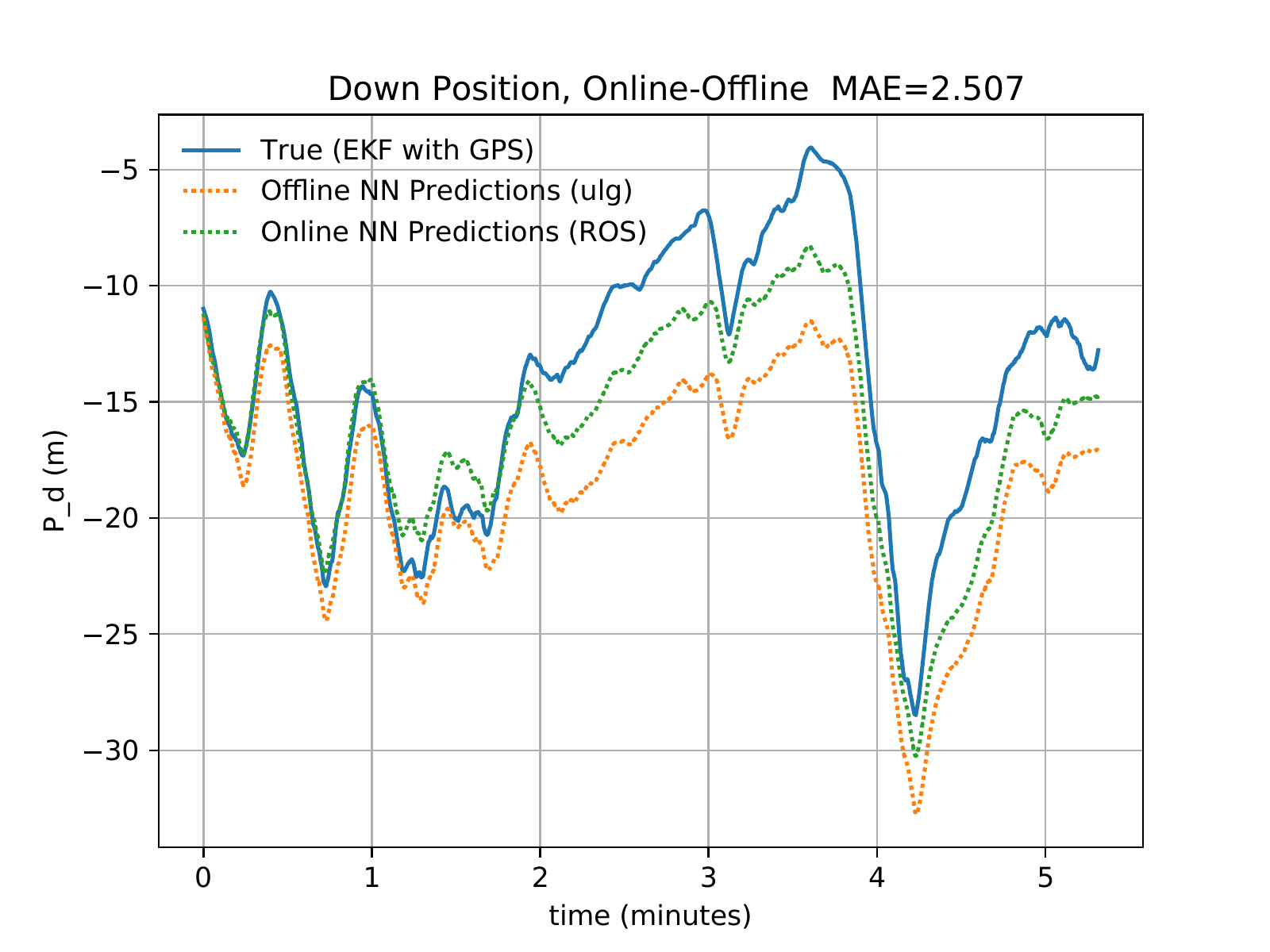}
	\caption{Validation ground trip, online	vs. offline predictions, north position}
	\label{fig:north_position}	
\end{figure}

\begin{figure}[h]
	\centering
	\includegraphics[width=1\linewidth]{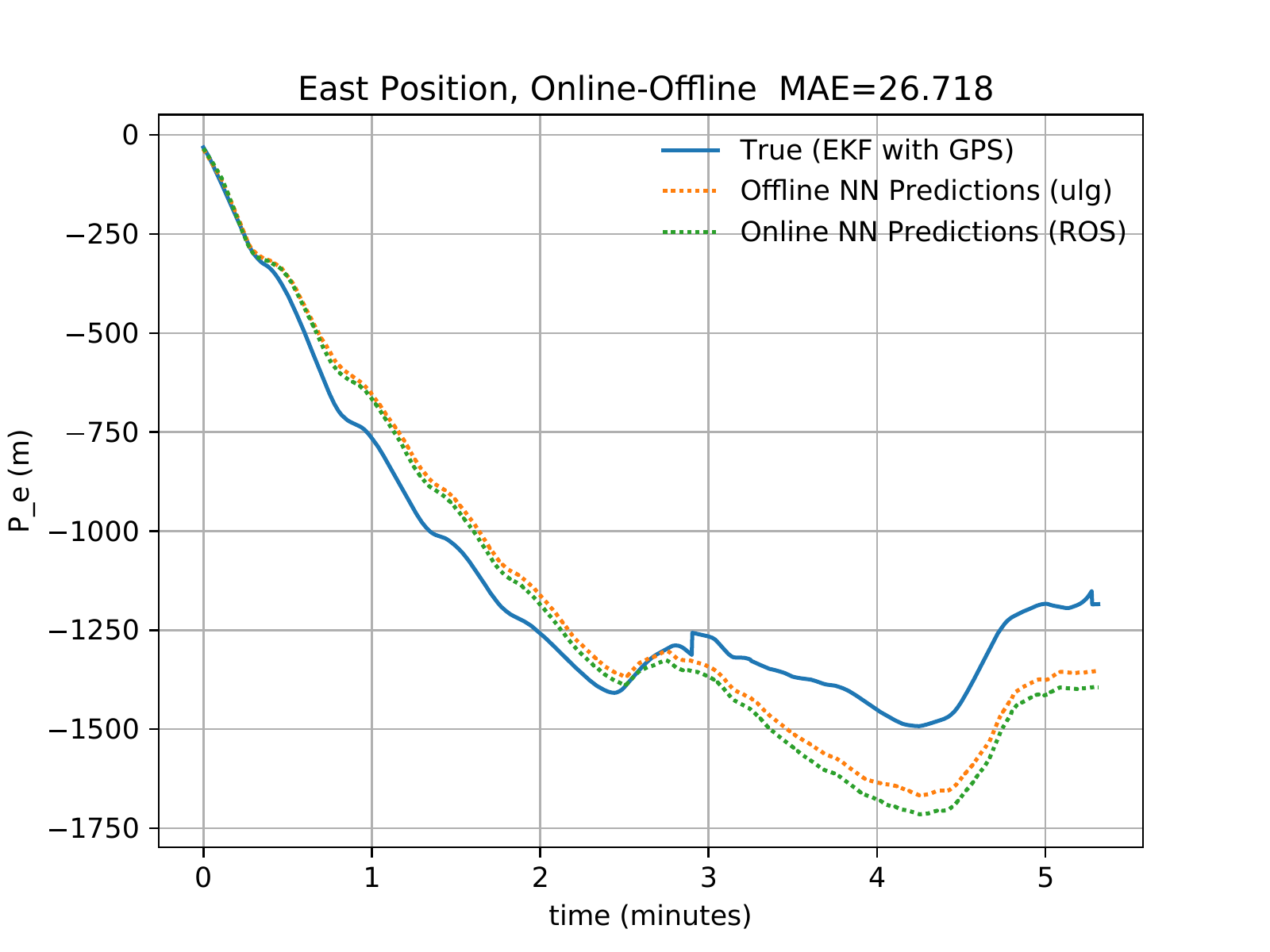}
	\caption{Validation ground trip, online	vs. offline predictions, east position}
	\label{fig:east_position}	
\end{figure}

\begin{figure}[h]
	\centering
	\includegraphics[width=1\linewidth]{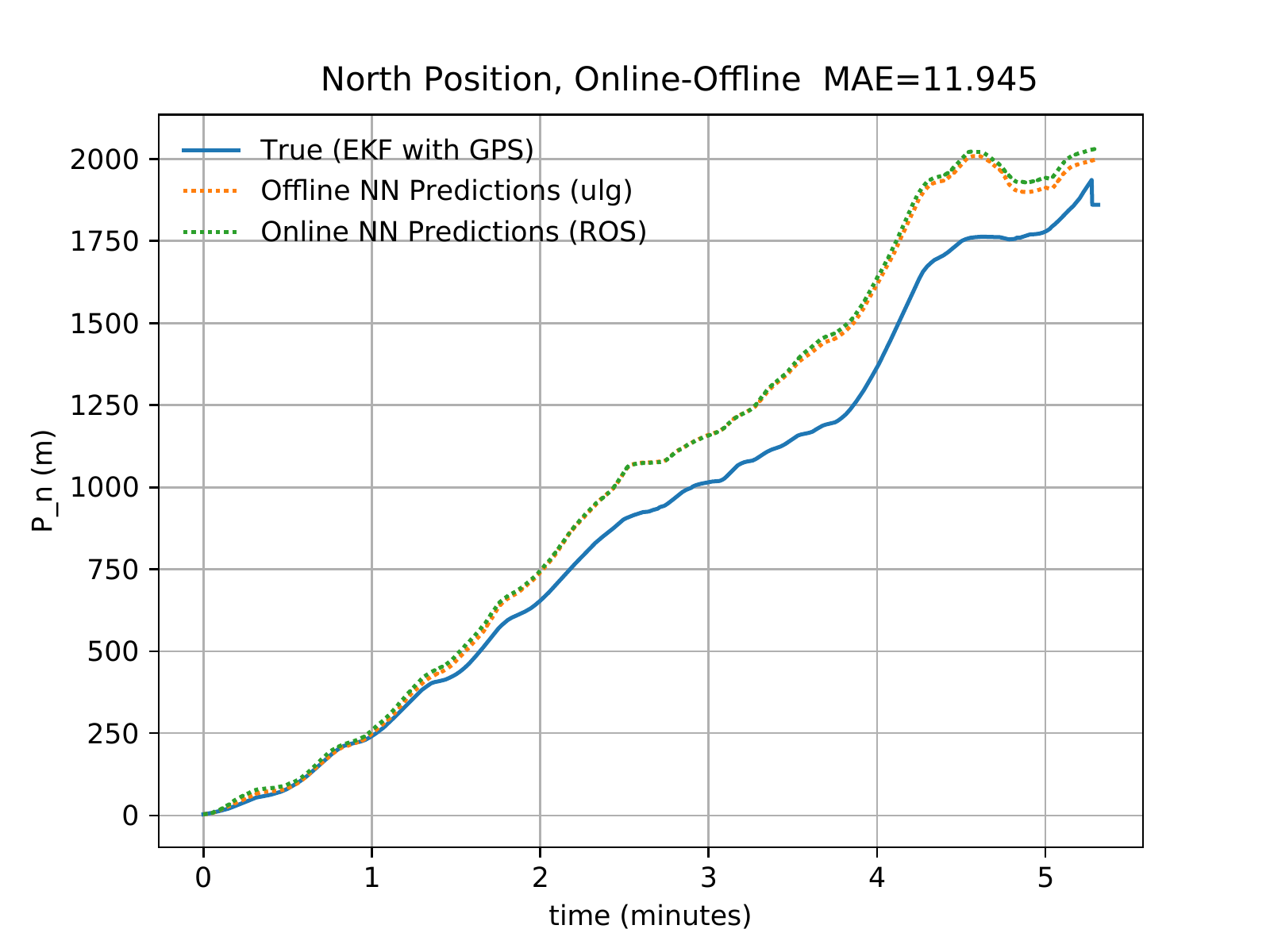}
	\caption{Validation ground trip, online	vs. offline predictions, down position}
	\label{fig:down_position}	
\end{figure}

Several complications arise when edge computing is performed. Aside
from accuracy loss resulting from the inevitable network size reduction,
other issues like timing, proper preprocessing, data validity, and
latency also affect the NN predictions. Most of these complications
affect the features consumed by the network. For example, the network
makes a prediction every 200 milliseconds; timing this loop on a full
operating system -rather than a real-time OS- is hard to guarantee.
Consequently, a \textpm 1 ms timing error is introduced, this affects
both averaging and differencing steps in preprocessing. Furthermore,
since sensors operate at different rates, multi-threading is needed
to capture every sensor's measurements as soon as they are published.
And to avoid data loss, these measurements should be stored in queues
before they are fed to the NN. But if these queues are allowed to
grow indefinitely, then latency in NN predictions will be noticed,
so a compromise should be made. Another issue is data validity; having
different functions writing to the features array can easily result
in data corruption due to race conditions. After addressing all these
issues, the NN predictions made in real-time are compared to those
made offline using the logged data. Figure \ref{fig:north_position},
Figure \ref{fig:east_position} and Figure \ref{fig:down_position}
show the online and offline position predictions made on a validation
trip.

\section{Summary and Conclusion}

The paper presented a Neural Network capable of predicting
a drone's 3D position and speed in the absence of GPS using the IMU,
Barometer and Magnetometer raw measurements. The developed network
was tested on several hardware targets, and can run with acceptable
speed on embedded devices like the Jetson Nano. Errors in 3D position
as low as 2.7 were achieved using the proposed system in 5-minutes
GPS-less drone flight using low-cost sensors. The system is well integrated
with the PX4 flight stack running on Pixhawk4 and can be used with
a quick setup. The system can also predict the attitude,
but the EKF is more reliable in such a task. The network is designed
to make inferences at 5 Hz, so it is slower than the EKF. This is
why the network's main role is to replace the GPS, not the entire
EKF. The network is ideally used with Quadrotors running on ``Auto''
flight mode because most of the training data have these characteristics.
Transfer learning is also used, and the system was tested in real-time
with a ground host vehicle. The network's errors associated with manual
maneuvers or less popular host vehicles can be reduced by collecting
more training data focusing on these conditions. Navigation accuracy
also increases when a larger network and window size are used. But this
comes at the expense of both training and inference times. The median
positioning error on the validation set is five times larger than
that of the training set. This signals the need for proper regularization
to help the network generalize better.

\bibliographystyle{elsarticle-num} 
\bibliography{references_database}

\end{document}